# Subroutines to Simulate Fission Neutrons for Monte Carlo Transport Codes

J. P. Lestone
Los Alamos National Laboratory, Los Alamos, New Mexico, USA

## Abstract

Fortran subroutines have been written to simulate the production of fission neutrons from the spontaneous fission of $^{252}$Cf and $^{240}$Pu, and from the thermal neutron induced fission of $^{239}$Pu and $^{235}$U. The names of these four subroutines are *getnv252*, *getnv240*, *getnv239*, and *getnv235*, respectively. These subroutines reproduce measured first, second, and third moments of the neutron multiplicity distributions, measured neutron-fission correlation data for the spontaneous fission of $^{252}$Cf, and measured neutron-neutron correlation data for both the spontaneous fission of $^{252}$Cf and the thermal neutron induced fission of $^{235}$U. The codes presented here can be used to study the possible uses of neutron-neutron correlations in the area of transparency measurements and the uses of neutron-neutron correlations in coincidence neutron imaging.

## Introduction

The most studied fissioning system is the spontaneous fission of $^{252}$Cf. Detailed data exist for (1) the fission fragment mass distribution, (2) the average total kinetic energy and the variance of the total kinetic energy as functions of fragment mass, (3) the average neutron multiplicity as a function of fragment mass, (4) the variation of neutron multiplicity with total kinetic energy of the fragments, (5) the neutron energy spectra as a function of fragment mass, (6) the total neutron energy spectrum, (7) the neutron-fission correlation, (8) neutron-neutron correlations, and (9) the first, second, and third moments of the total neutron multiplicity distribution. If one's aim is to develop codes to simulate fission neutron production to be used by Monte Carlo neutron transport codes, then it is logical to first start with the spontaneous fission of $^{252}$Cf.

The experimental inputs used to simulate the fission neutrons from the spontaneous fission of $^{252}$Cf are the fission fragment mass distribution, the average total kinetic energy and the variance of the total kinetic energy as functions of fragment mass, the average neutron multiplicity as a function of fragment mass, the variation of neutron multiplicity with total kinetic energy of the fragments, and the neutron energy spectra as a function of fragment mass in the reference frame of the emitting fragments. It is well established that most fission neutrons are evaporated from fully accelerated fission fragments. The origin of the remaining ~10% of fission neutrons are due to a combination of neutrons emitted during the fission process before the fissioning system breaks into two separate fragments and neutrons emitted by the fission fragment before they are fully accelerated. The precise nature of these neutrons is, however, poorly understood. In the modeling presented here, it is assumed that all fission neutrons are evaporated isotopically in the rest frame of fully accelerated fission fragments.

To accurately reproduce well-established first, second, and third moments of the $^{252}$Cf neutron multiplicity distribution and the mean energy of $^{252}$Cf fission neutrons, it was necessary to make minor modifications to the average neutron multiplicities as a function of fragment mass, the variation of neutron multiplicities with fragment kinetic energy, and the neutron energy spectra in the reference frame of the emitting fragments. With these modifications the simulations presented here also reproduce measured neutron-fission correlations. One property of the fission neutrons that has not been measured is the standard deviation of neutron multiplicities per fragment for fixed mass splits and fixed total kinetic energies. These unmeasured standard deviations affect the neutron-neutron correlations. If the standard deviation of the neutron multiplicities at fixed masses and total kinetic energies is assumed to be zero, then the present model calculations fail to reproduce observed $^{252}$Cf neutron-neutron correlations. If a standard deviation of 0.75 is assumed for the neutron multiplicities per fragment, for all mass splits, and for all total kinetic energies, then the present model provides an adequate fit to measured neutron-neutron correlations.

Similar data were used to develop models of the neutron production in the spontaneous fission of $^{240}$Pu and the thermal neutron induced fission of $^{239}$Pu and $^{235}$U. As with the $^{252}$Cf case, minor changes to the inputted experimental data were needed before adequate agreement was obtained between the model calculations and the measured first, second, and third moments of the neutron multiplicity distributions and the mean energies of the corresponding fission neutrons. If a standard deviation of 0.75 is assumed for the neutron multiplicities per fragment, for all mass splits, and for all total kinetic energies, then the present simulations also provide a good description of the available neutron-neutron correlation data for thermal neutron induced fission of $^{235}$U.

## The Fortran Subroutines

Four Fortran subroutines have been written to simulate fission-neutron production from the spontaneous fission of $^{252}$Cf and $^{240}$Pu and the thermal neutron induced fission of $^{239}$Pu and $^{235}$U. These subroutines have the names, *getnv252*, *getnv240*, *getnv239*, and *getnv235*, respectively, and can be called using the Fortran statement “call getnvxxx(itot,nv,ffv)”, where itot, nv, and vff are declared using the statements “Integer itot” and “Real*8 nv(1:10,1:3), vff(1:2,1:3)”. On return to the calling program, itot contains the number of neutrons from a random fission event, and nv(i,j) contains the velocities of the i=1 to itot fission neutrons. The indices j=1, 2, and 3 represent the x, y, and z directions. For example nv(3,2) would be the y component of the velocity of the third fission neutron. vff(i,j) contains the velocities of the two fission fragments. In this case the indices i=1 and 2 are for the light and heavy fission fragments. All velocities are in units of 0.98-cm/ns. This is a convenient choice because the energy of neutrons in MeV can be obtained using the expression $E=v^2/2$.

It is convenient if the computer time required to generate source neutrons is much smaller than the time required to track these neutrons using Monte Carlo transport codes like MCNP4A. One million calls to the *getnvxxx* subroutines take less than 30 seconds on a 266 MHz computer. This corresponds to $10^6$ fission neutrons in less than 10 seconds. This is substantially faster than the time required for MCNP4A to track the corresponding number of source neutrons through a system of modest complexity. If the need arose for faster source neutron production, then the *getnvxxx* subroutines could possibly be altered to run ~50% faster.

## Simulation of Neutrons from the Spontaneous Fission of $^{252}$Cf

To simulate the neutron emission from the spontaneous fission of $^{252}$Cf, it is necessary to first simulate the distribution of fission fragment velocities. This is done by randomly picking a mass split from the mass yield curve. For each mass split the average total kinetic energy (TKE) of the fission fragments before neutron evaporation and the corresponding variance are well known. Figures 1, 2, and 3 show the mass yield curves for the spontaneous fission of $^{252}$Cf[1] and the corresponding average and standard deviation of the TKE as a function of the light fragment mass[1]. The important components in simulating the neutron multiplicity distribution are the mass yield curve, the average neutron multiplicity per fragment[1], and its dependence on the TKE[1,2] (see figures 1, 4, and 5). Measuring the dependence of the neutron multiplicity per fragment on TKE is problematic and the results from different studies are in disagreement (see figure 5).

If the experimental results shown in figures 1, 3, and 4, and data of Budtz-Jørgensen and Knitter shown in figure 5 are used to simulate the neutron multiplicity distribution for $^{252}$Cf, then the resulting first, second, and third moments are 3.67, 11.89, and 33.19. These are similar to, but not in excellent agreement with, the corresponding measured values of 3.76, 11.96, and 31.81[3]. This discrepancy can be overcome by scaling the results shown in figure 4 by 1.02 and by scaling the results of Budtz-Jørgensen and Knitter shown in figure 5 by 0.87. The resulting simulated first, second, and third moments are then 3.76, 11.98, and 31.72. The need for 1.02 scaling of the neutron multiplicity data shown in figure 4 could be simply associated with errors in scanning the data from photocopied figures from the original paper[1]. The 0.87 scaling of the linear dependence of the neutron multiplicity per fragment on TKE, of Budtz-Jørgensen and Knitter, could be associated with either the previously discussed problematic nature of measuring this quantity, the fact that the TKE distribution for a fixed mass split is not purely Gaussian, or non-linear components in the dependence of neutron multiplicity per fragment on TKE. The two scaling factors discussed above are used in all the following simulations of neutrons from $^{252}$Cf. Figure 6 compares the experimentally deduced neutron multiplicity distribution from the spontaneous fission of $^{252}$Cf[3] to the corresponding simulated distribution.

To be able to simulate neutron-neutron correlations, it is necessary to first describe the energy (velocity) spectra of neutrons relative to the emitting fission fragments. The energy spectrum of neutrons evaporated from hot nuclei can be expressed as

$$P(E)\,dE \propto E^{\lambda} \exp(-E/T)\,. \qquad (1)$$

Figures 7 and 8 show the measurements of Budtz-Jørgensen and Knitter[1] for $\lambda$ and $T$ as a function of fragment mass. These $\lambda$ and $T$, in conjunction with the fission-fragment velocities determined using the results in figures 1, 2, and 3, and with the assumption that all neutrons are evaporated isotopically from fully accelerated fragments, lead to a simulated mean neutron energy of 2.26 MeV. This is less than the well-established mean energy for $^{252}$Cf fission neutrons of 2.34 MeV[4,5,6]. This is not unexpected since the temperature of the $^{252}$Cf total neutron energy spectrum in the laboratory frame measured by Budtz-Jørgensen and Knitter[1] is also significantly less than the corresponding values measured by Bowman et al.[4] and Meadows[5]. This problem can be solved by scaling the temperatures shown in figure 8 by 1.05. This leads to a simulated $^{252}Cf$ mean neutron energy of 2.34 MeV. Figure 9 shows a comparison of the MCNP4A recommended $^{252}Cf$ neutron-energy spectrum with the corresponding simulated spectrum obtained using the subroutine *getnv252*. Figure 10 shows the measured $^{252}Cf$ neutron-fission correlation with a neutron threshold of 0.5 MeV[4] and the corresponding simulated correlation. Figures 11 and 12 compare measured $^{252}Cf$ neutron velocity spectra as a function of angle to the light fission fragments[4] to simulated results obtained using the subroutine *getnv252*.

The standard deviation of the neutron multiplicity per fragment for fixed mass split and fixed TKE has not been measured but is required to simulate neutron-neutron correlations. To illustrate this, consider the case where at a given mass split and TKE the total multiplicity is 2, and the average multiplicity per fragment is 1. If one neutron is also emitted per fragment then the corresponding neutron-neutron correlation will strongly favor antiparallel neutrons. If, however, 1 neutron per fragment is just a probable as 2 neutrons from the light fragment with 0 neutrons from the heavy fragment, and 2 neutrons from the heavy fragment with 0 neutrons from the light fragment, then the neutron-neutron correlation will favor parallel neutrons. Notice that both of the above examples have the same total multiplicity per fission event and the same average multiplicity per light and heavy fragments but lead to very different neutron-neutron correlations. To incorporate these effects into the neutron simulations presented here, the standard deviation of all neutron multiplicities per fragment, at all fragment masses and TKEs, is assumed to be a constant, $\sigma_N$. This value is used in the following way. Using the data (scaled versions in some cases) shown in figures 1-5, the fission mass split, TKE, and multiplicity per light and heavy fragments is randomly chosen. A number is then chosen from a Gaussian distribution with standard deviation $\sigma_N$. This number is subtracted from the multiplicity per light fragment and added to the multiplicity per heavy fragment. The inclusion of this procedure leaves the simulated neutron multiplicity distributions unchanged but has a strong influence on the simulated neutron-neutron correlations.

Figure 13 shows the measured $^{252}Cf$ neutron-neutron correlation of Pringle and Brooks[7]. The dashed line shows the simulated $^{252}Cf$ neutron-neutron correlation for all neutron energies with $\sigma_N$ set to zero. The thin solid line shows the corresponding correlation but with the experimental neutron detection efficiency of Pringle and Brooks taken into account. This simulated neutron-neutron correlation is in disagreement with the experimental data. The thick solid line shows the simulated $^{252}Cf$ neutron-neutron correlation with $\sigma_N$=0.75.

**Simulation of Neutrons from the Thermal Neutron Induced Fission of $^{235}U$**

Figures 14, 15, and 16 show the mass yield curve for thermal neutron induced fission of $^{235}U$[8] and the corresponding average and standard deviation of the TKE as a function of the light fragment mass[9]. Figures 17 and 18 show the average neutron multiplicity per fragment[10] and its dependence on the TKE[11]. If the experimental results shown in figures 14, 16, and 17, and shifted $^{239}Pu(n_{th},f)$ data of Basova et al.[11] shown in figure 18 are used to simulate the neutron multiplicity distribution for $^{235}U(n_{th},f)$, then the resulting first, second, and third moments are 2.52, 5.81, and 11.83. These are in disagreement with the corresponding measured values of 2.41, 4.63, and 6.86[3]. This discrepancy can be overcome by scaling the results shown in figure 17 by 0.97 and by scaling the shifted results of Basova et al.[11] shown in figure 18 by 0.76. The resulting simulated first, second, and third moments are then 2.40, 4.64, and 6.85. The data in figures 14-18, along with the above-discussed two scaling factors, were used in all the following simulations of neutrons from the thermal neutron induced fission of $^{235}U$. Figure 19 compares the experimentally deduced neutron multiplicity distribution for $^{235}U(n_{th},f)$[3] to the corresponding simulated distribution.

The neutron energy spectra in the reference frame of all $^{235}U(n_{th},f)$ fission fragments is assumed to be of the form given in equation 1 with $\lambda$=½. If $T$ is chosen to be 0.84 MeV, then the simulated mean energy of $^{235}U(n_{th},f)$ fission neutrons is in agreement with the known value of 2.03 MeV[6]. Figure 20 shows a comparison of the MCNP4A recommended $^{235}U(n_{th},f)$ neutron-energy spectrum[6] with the corresponding simulated spectrum obtained using the subroutine *getnv235*. Figures 21, 22, and 23 show measured $^{235}U(n_{th},f)$ neutron-neutron correlations[12]. The thick solid lines show the corresponding simulated neutron-neutron correlations obtained using $\sigma_N$=0.75. Notice that $\sigma_N$=0.75 leads to a satisfactory reproduction of both the available $^{252}Cf$ and $^{235}U(n_{th},f)$ neutron-neutron correlations. This suggests that, in the absence of data for other isotopes, the neutron-neutron correlations for other fissioning systems can be predicted using $\sigma_N$=0.75.

**Simulation of Neutrons from the Thermal Neutron Induced Fission of $^{239}$Pu**

Figures 24, 25, and 26 show the mass yield curve for the fission of $^{240}$Pu and the corresponding average and standard deviation of the TKE as a function of the light fragment mass[13]. Figures 27 and 28 show the average neutron multiplicity per fragment and its dependence on the TKE[11]. If the experimental results shown in figures 24, 26, 27, and 28 are used to simulate the neutron multiplicity distribution for $^{239}$Pu($n_{th}$,*f*), then the resulting first, second, and third moments are 2.92, 7.99, and 19.45. These are in disagreement with the corresponding measured values of 2.88, 6.77, and 12.63[3]. This discrepancy can be overcome by scaling the results shown in figure 27 by 0.99 and by scaling the results shown in figure 28 by 0.72. The resulting simulated first, second, and third moments are then 2.87, 6.80, and 12.60. The data in figures 24–28, along with the above discussed two scaling factors, were used in all the following simulations of neutrons from the thermal neutron induced fission of $^{239}$Pu. Figure 29 compares the experimentally deduced neutron multiplicity distribution for $^{239}$Pu($n_{th}$,*f*)[3] to the corresponding simulated distribution.

The neutron energy spectra in the reference frame of all $^{239}$Pu($n_{th}$,*f*) fission fragments is assumed to be of the form given in equation 1 with $\lambda$=½. If *T* is chosen to be 0.89 MeV then the simulated mean energy of $^{239}$Pu($n_{th}$,*f*) neutrons is in agreement with the known value of 2.11 MeV[6]. Figure 30 shows a comparison of the MCNP4A recommended $^{239}$Pu($n_{th}$,*f*) neutron-energy spectrum[6] with the corresponding simulated spectrum obtained using the subroutine *getnv239*. Figure 31 shows simulated $^{239}$Pu($n_{th}$,*f*) neutron-neutron correlations with various neutron energy thresholds. Based on the satisfactory reproduction of the $^{252}$Cf and $^{235}$U($n_{th}$,*f*) data, the value for $\sigma_N$ used was 0.75.

**Simulation of Neutrons from the Spontaneous Fission of $^{240}$Pu**

The mass yield curve for the spontaneous fission of $^{240}$Pu, the corresponding average and standard deviation of the TKE, and the dependence of the neutron multiplicity per fragment on TKE as a function of the light fragment mass, is shown in figures 24, 25, and 26. The assumed $^{240}$Pu spontaneous-fission average neutron multiplicity per fragment as a function of the fragment mass is shown in figure 32. These $^{240}$Pu spontaneous-fission average neutron multiplicities per fission fragment were estimated using the corresponding $^{239}$Pu($n_{th}$,*f*) values and the expression

$$\nu_{240}(A_{FF}) = \nu(A_{FF}) - \frac{\nu(A_{FF}) \times 0.723}{\nu(A_{FF}) + \nu(240 - A_{FF})} . \quad (2)$$

This equation reduces the neutron multiplicity for each mass split by 0.723 and reduces the multiplicity per fragment proportional to the $^{239}$Pu($n_{th}$,*f*) values. If the experimental results shown in figures 24, 26, 28, and 32 are used to simulate the neutron multiplicity distribution for the spontaneous fission of $^{240}$Pu, then the resulting first, second, and third moments are 2.26, 4.97, and 9.85. These are in disagreement with the corresponding measured values of 2.16, 3.83, and 5.34[3]. This discrepancy can be overcome by scaling the results shown in figure 32 by 0.98 and by scaling the results shown in figure 28 by 0.75. The resulting simulated first, second, and third moments are then 2.15, 3.84, and 5.30. The data in figures 24–26, 28, and 32, along with the above-discussed two scaling factors, were used in all the following simulations of neutrons from the spontaneous fission of $^{240}$Pu. Figure 33 compares the experimentally deduced neutron multiplicity distribution for the spontaneous fission $^{240}$Pu[3] to the corresponding simulated distribution.

The neutron energy spectra, in the reference frame of all $^{240}$Pu spontaneous fission fragments, is assumed to be of the form given in equation 1 with $\lambda$=½. If *T* is chosen to be 0.80 MeV, then the simulated mean energy of spontaneous $^{240}$Pu neutrons is in agreement with the known value of 1.97 MeV[6]. Figure 34 shows a comparison of the MCNP4A recommended $^{240}$Pu neutron-energy spectrum[6] with the corresponding simulated spectrum obtained using the subroutine *getnv240*. Figure 35 shows simulated neutron-neutron correlations for the spontaneous fission of $^{240}$Pu with various neutron energy thresholds. Based on the satisfactory reproduction of the $^{252}$Cf and $^{235}$U($n_{th}$,*f*) data, the value for $\sigma_N$ was 0.75.

**Summary**

Four Fortran subroutines have been written to simulate the prompt fission neutrons from the spontaneous fission of $^{252}$Cf and $^{240}$Pu, and the thermal neutron induced fission of $^{239}$Pu and $^{235}$U. These subroutines reproduce the experimentally known first, second, and third moments of the neutron multiplicity distributions to better than ~0.5%. They also produce total neutron energy spectra in good agreement with the MCNP4A[6] recommended spectra. The available $^{252}$Cf and $^{235}$U($n_{th}$,*f*) neutron-neutron correlation data are satisfactorily reproduced with $\sigma_N$=0.75. Based on this satisfactory reproduction of the $^{252}$Cf and $^{235}$U($n_{th}$,*f*) data the same value for $\sigma_N$ is used in simulating the $^{240}$Pu and $^{239}$Pu($n_{th}$,*f*) neutron-neutron correlations.

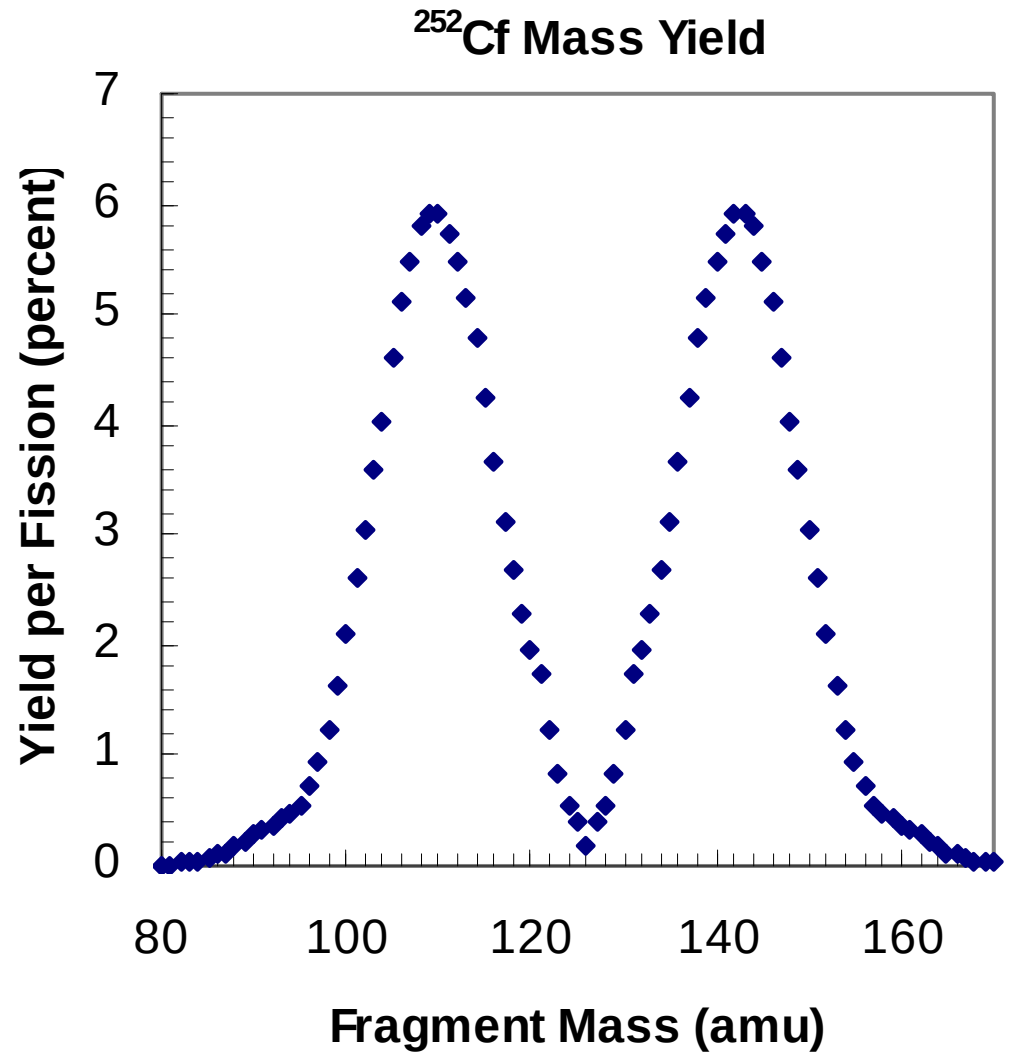

*Figure 1. ${}^{252}$Cf spontaneous fission mass yield*[1]*.*

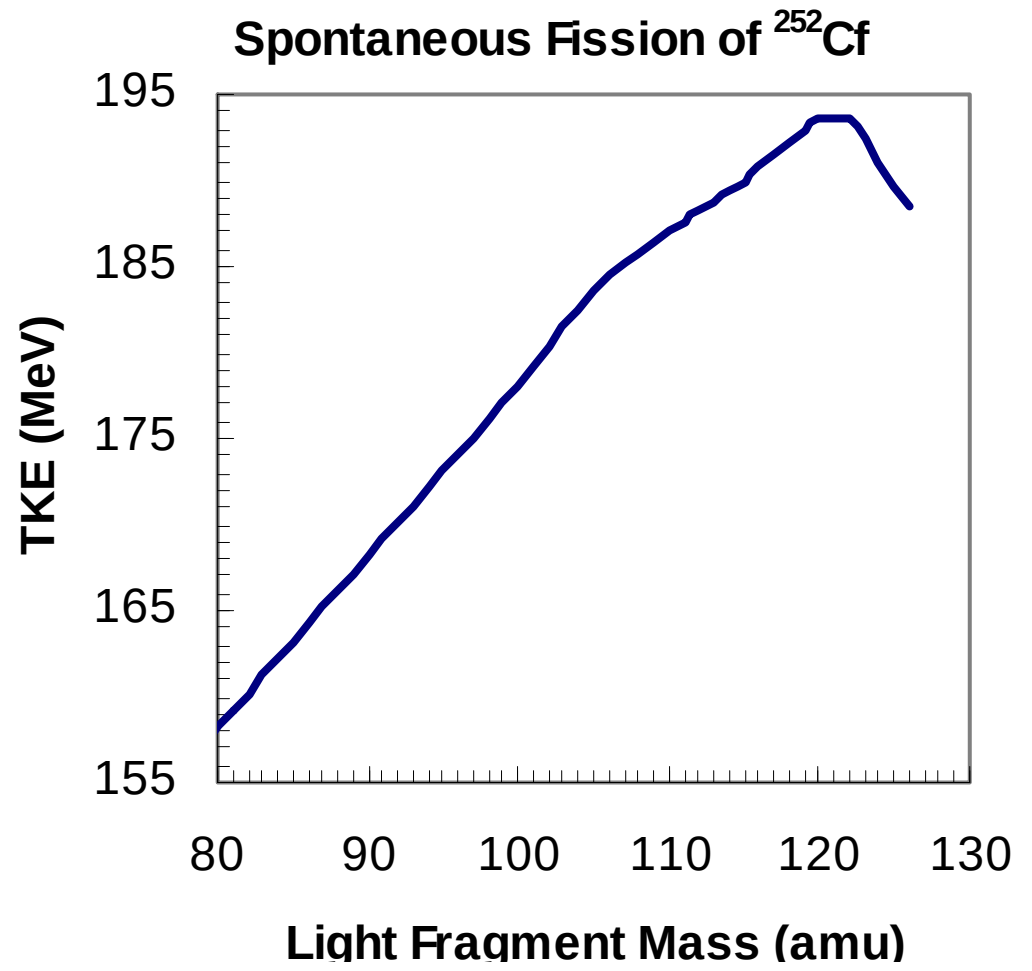

*Figure 2. Pre-neutron emission, total kinetic energy (TKE) release in the spontaneous fission of ${}^{252}$Cf as a function of the mass of the light fission fragment*[1]*.*

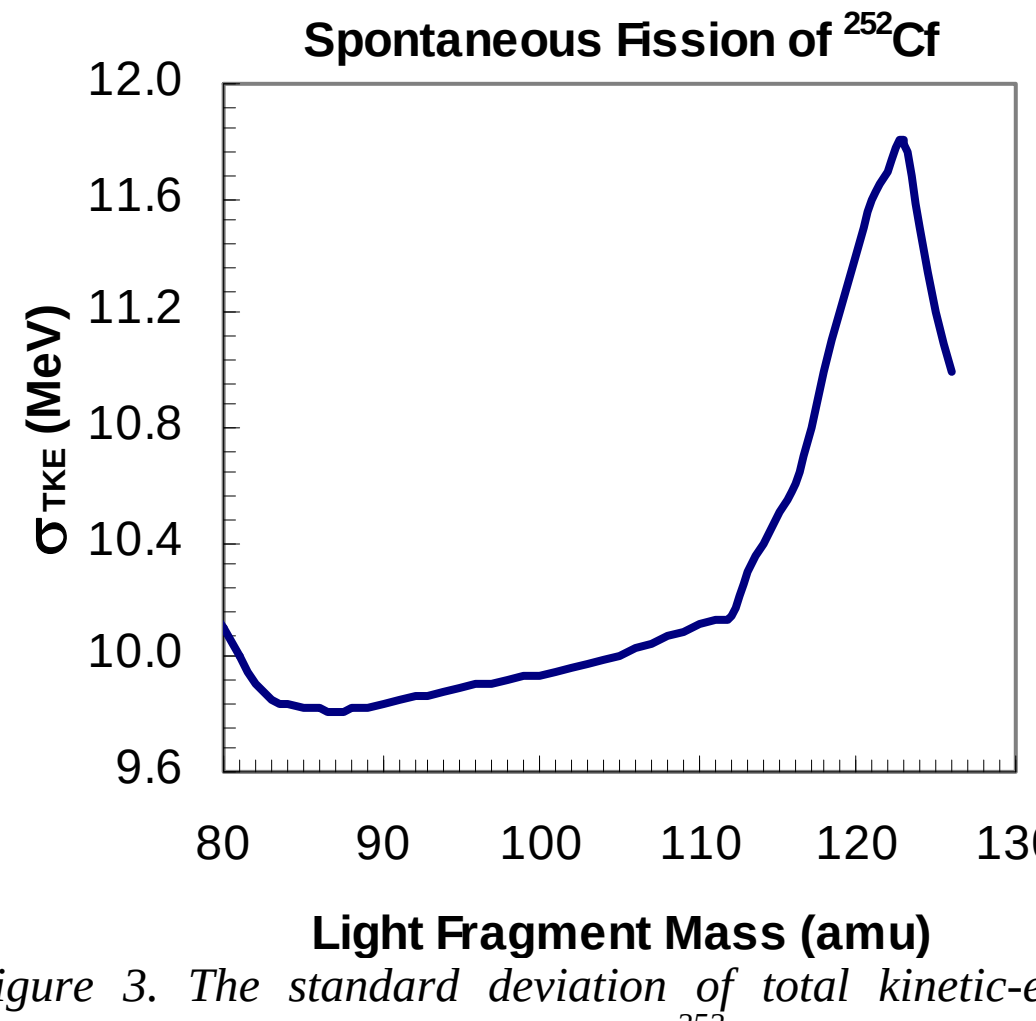

*Figure 3. The standard deviation of total kinetic-energy release in the spontaneous fission of ${}^{252}$Cf as a function of the mass of the light fission fragment*[1]*.*

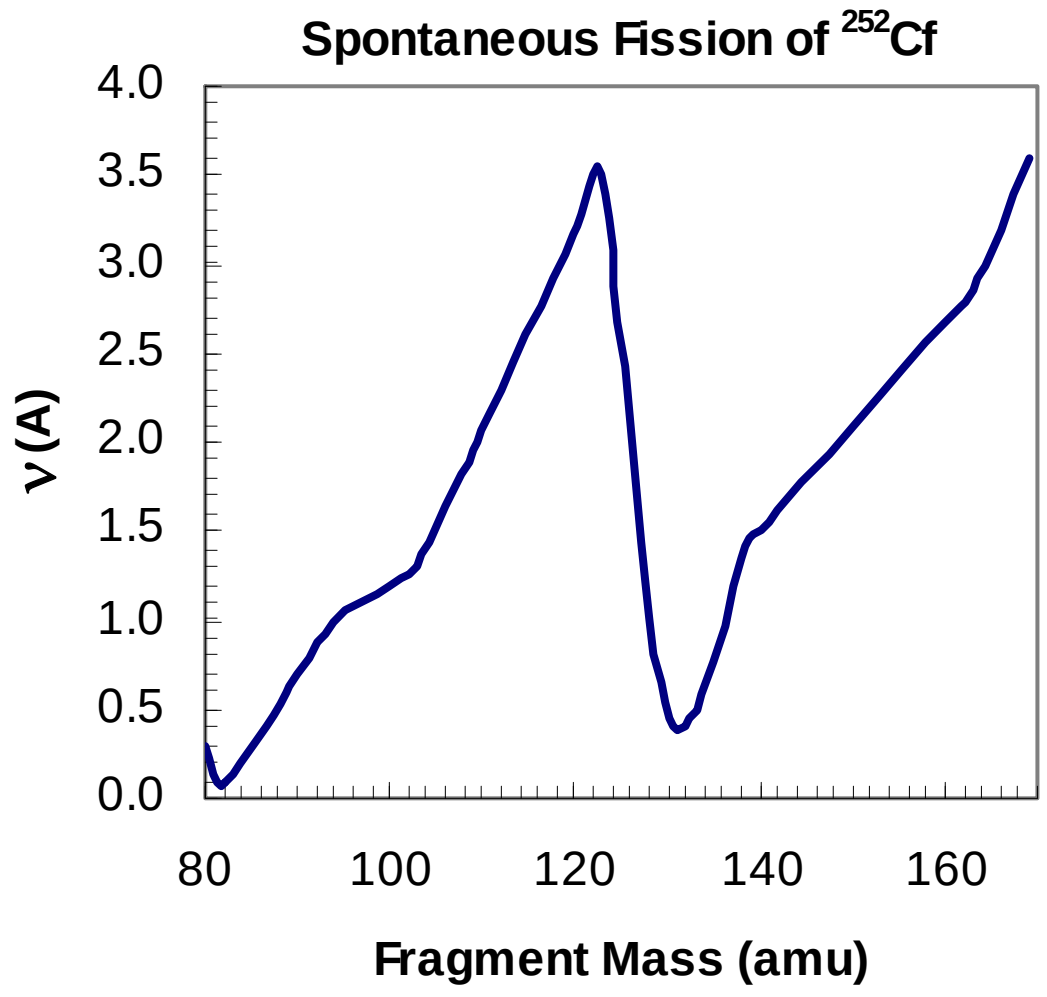

*Figure 4. The neutron multiplicity per fission fragment for the spontaneous fission of ${}^{252}$Cf*[1]*.*

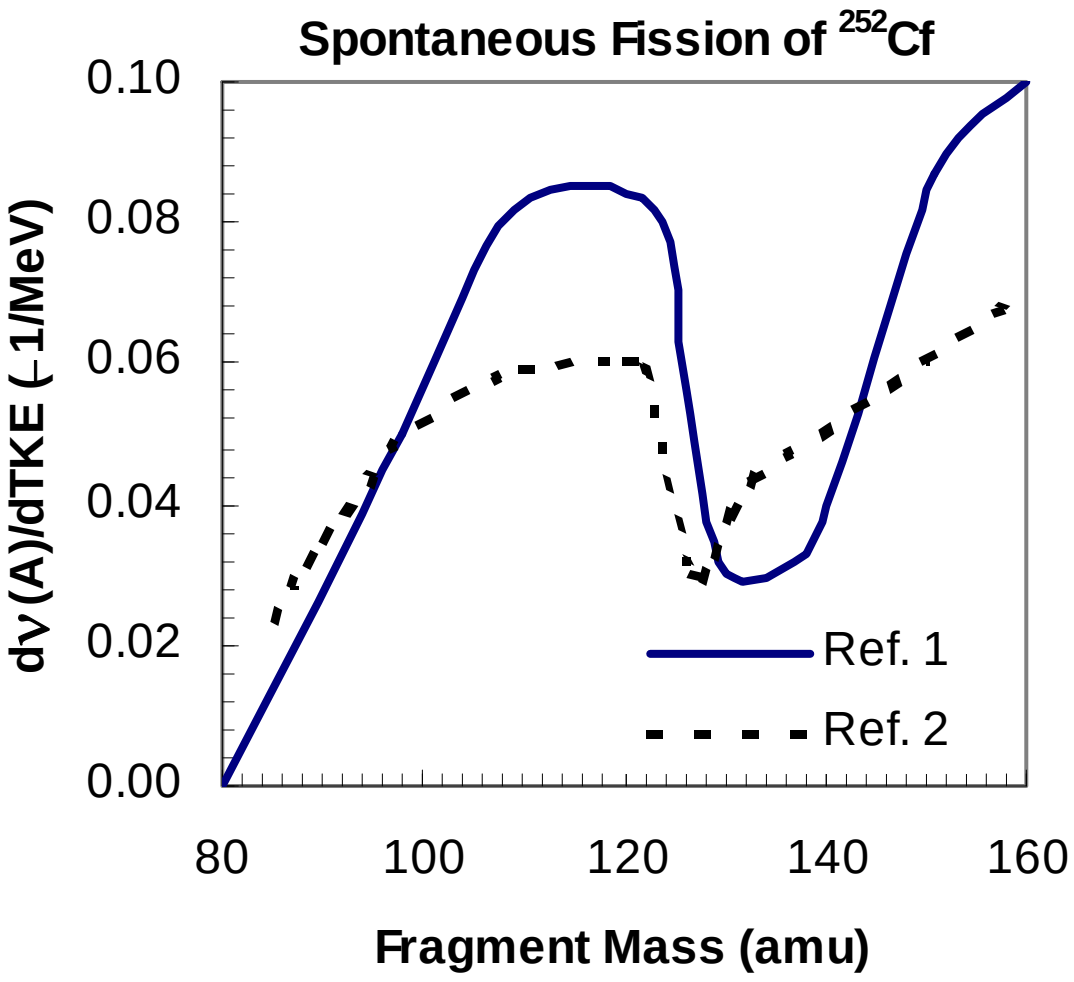

*Figure 5. The dependence of the multiplicity per fragment on the total kinetic energy released in the spontaneous fission of ${}^{252}$Cf.*

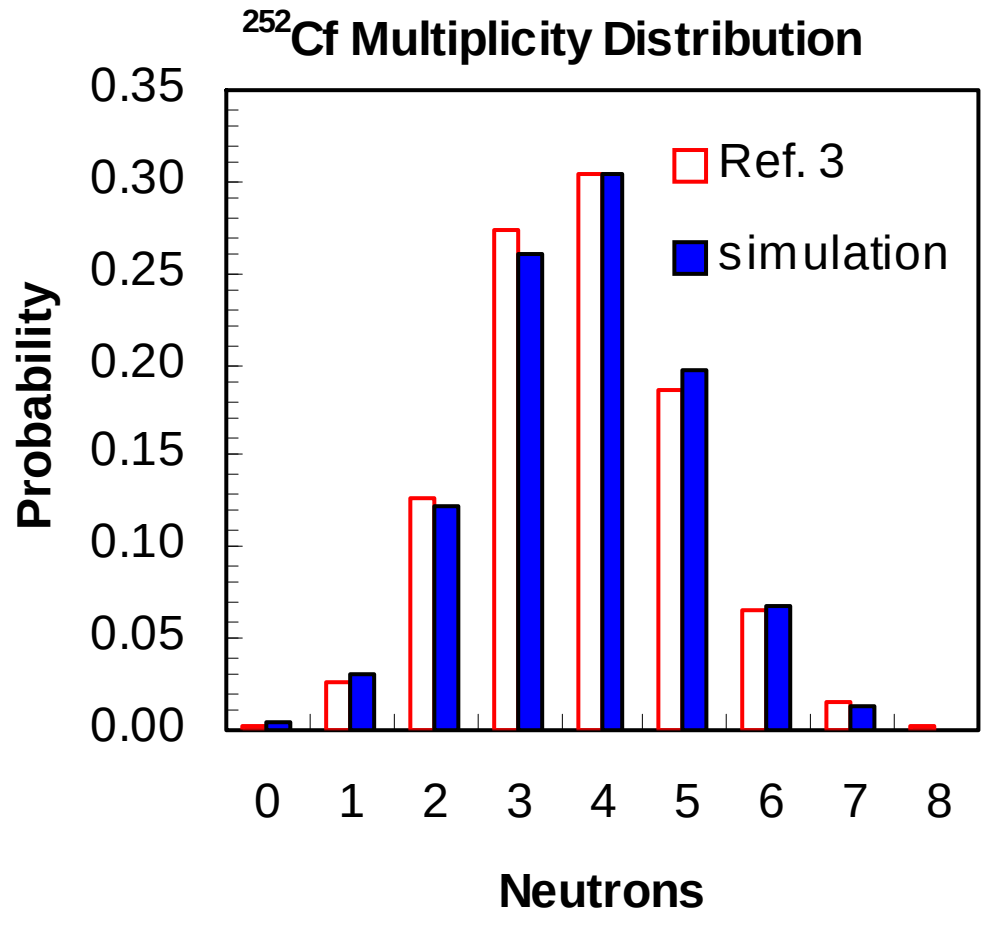

*Figure 6. Comparison of the experimentally deduced ${}^{252}$Cf spontaneous fission neutron multiplicity distribution of Boldeman and Hines*[3] *to the corresponding simulated distribution.*

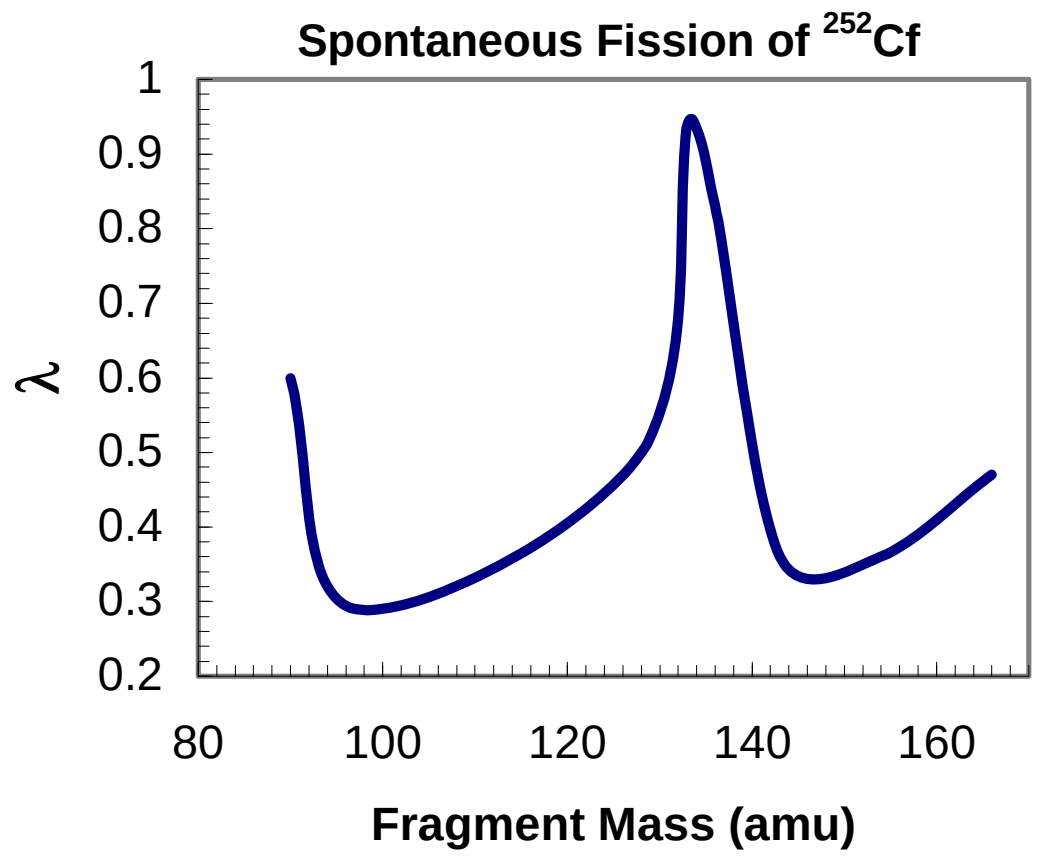

*Figure 7. The $^{252}Cf$ spontaneous fission neutron spectra $\lambda$ values as a function of fragment mass[1].*

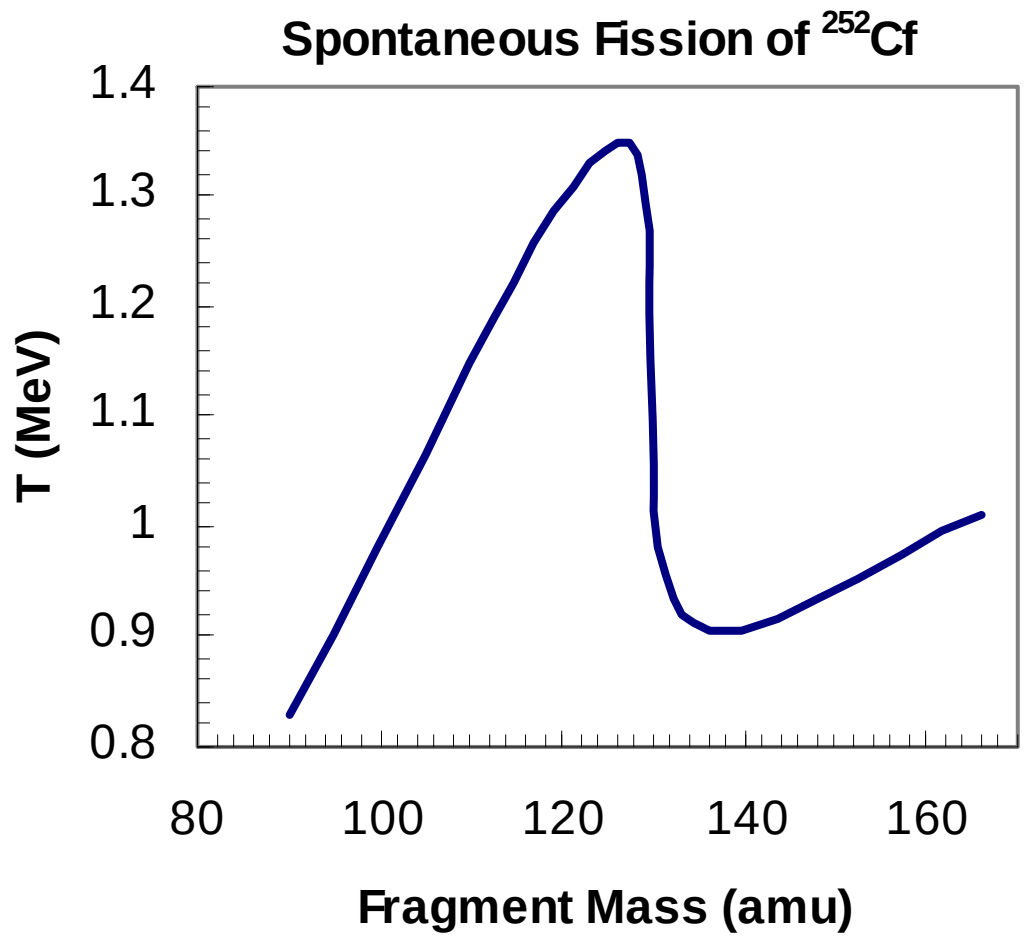

*Figure 8. The $^{252}Cf$ spontaneous fission neutron spectra temperatures, T, as a function of fragment mass[1].*

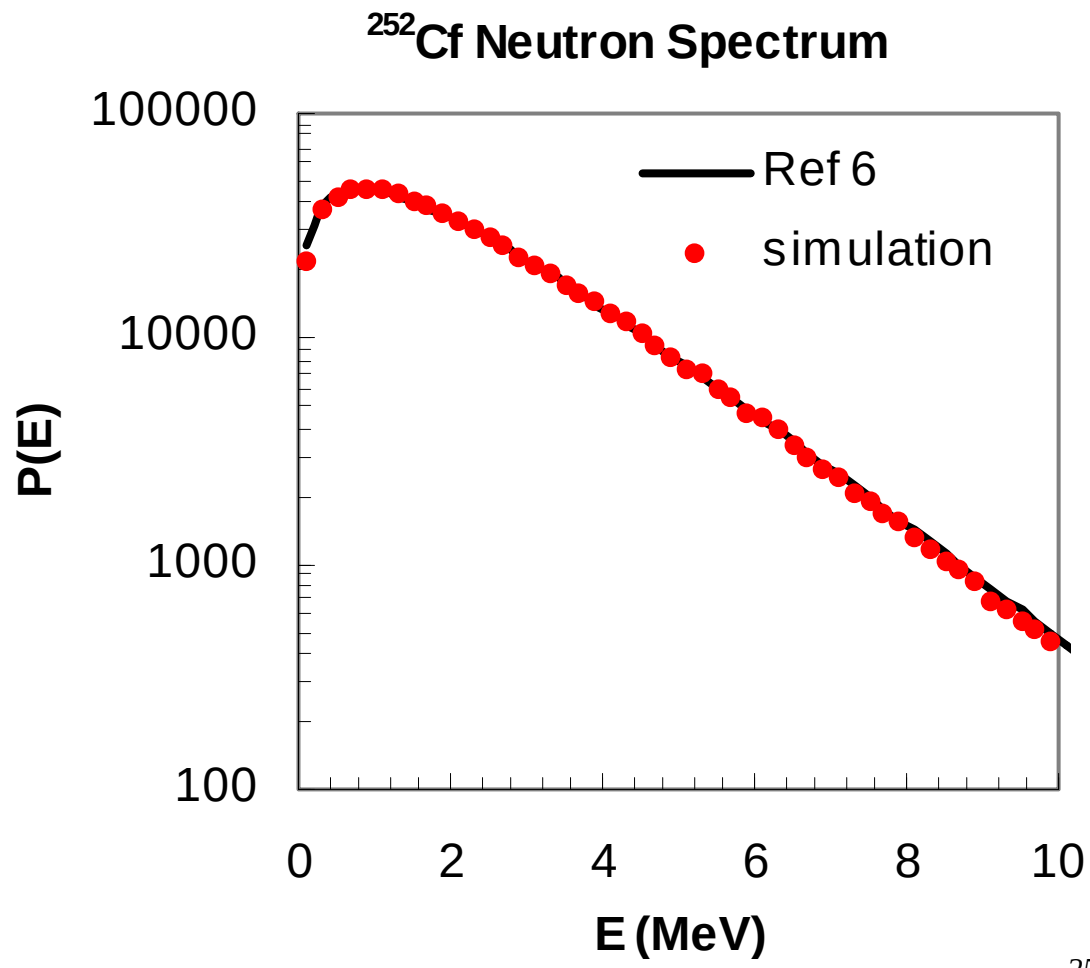

*Figure 9. Comparison of the MCNP4A recommended $^{252}Cf$ spontaneous fission neutron-energy spectrum to the corresponding simulated distribution.*

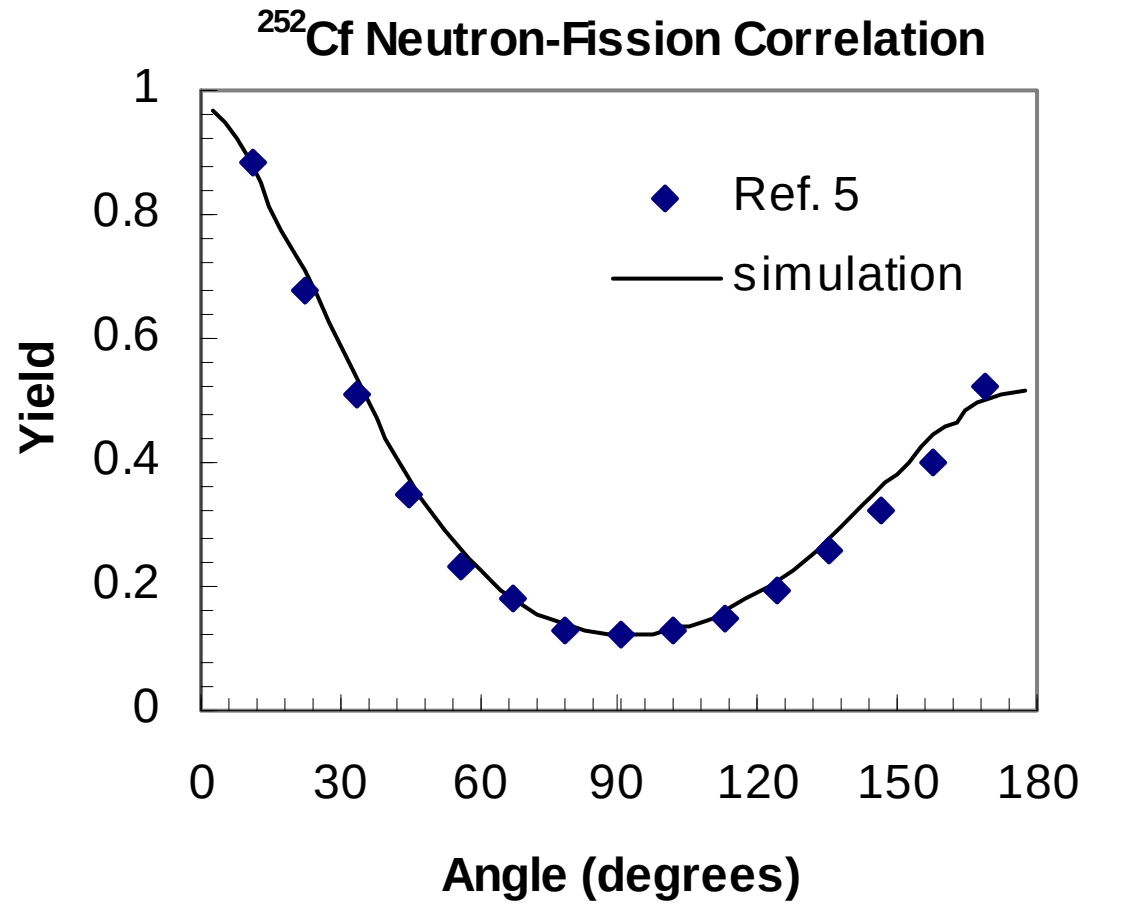

*Figure 10. Comparison of the measured $^{252}Cf$ neutron-fission correlation with a neutron threshold of 0.5 MeV[4] with the corresponding simulated correlation.*

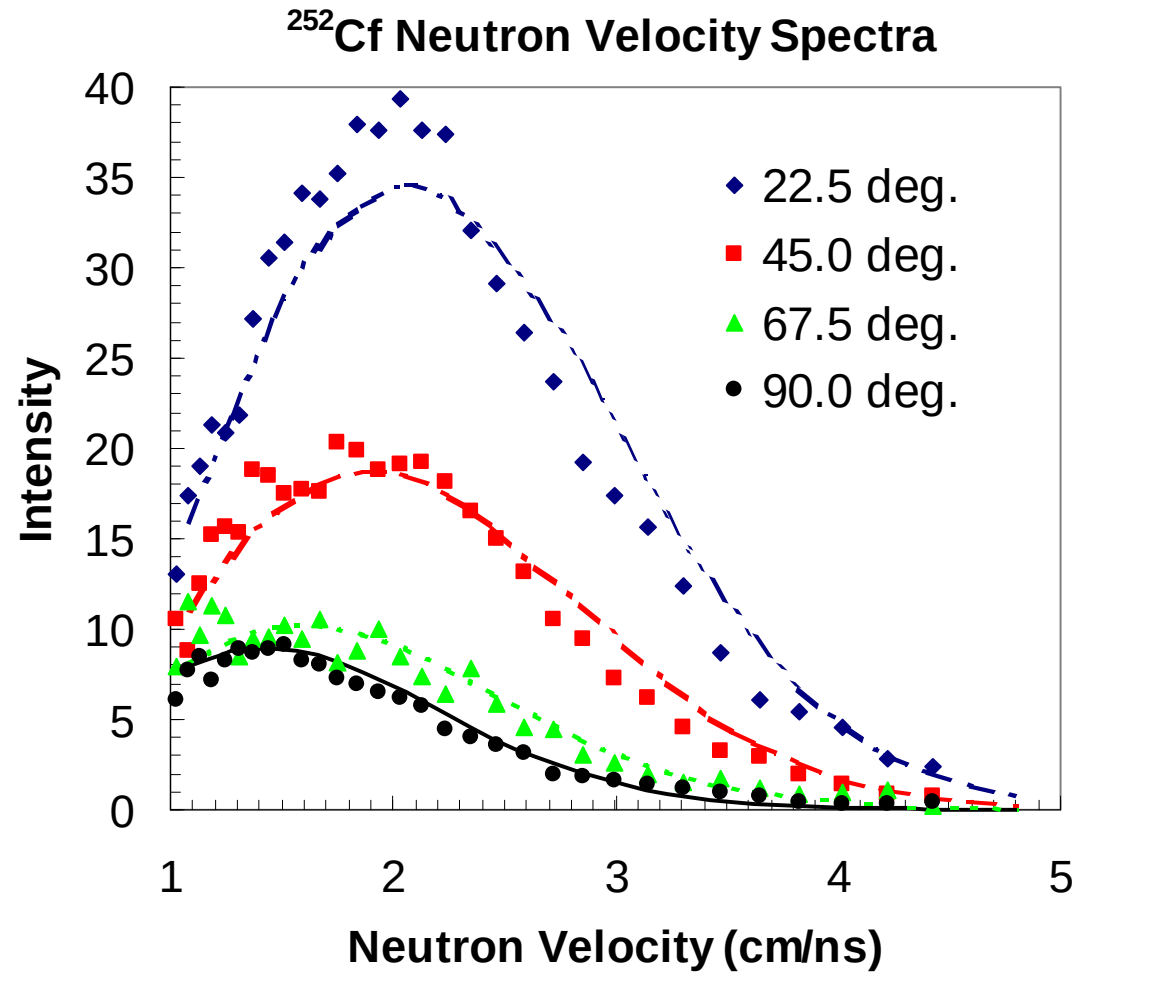

*Figure 11. The symbols are measured $^{252}Cf$ neutron-velocity spectra[4] at various angles between the neutrons and the light fission fragments $\leq 90°$. The smooth curves show the corresponding simulated spectra.*

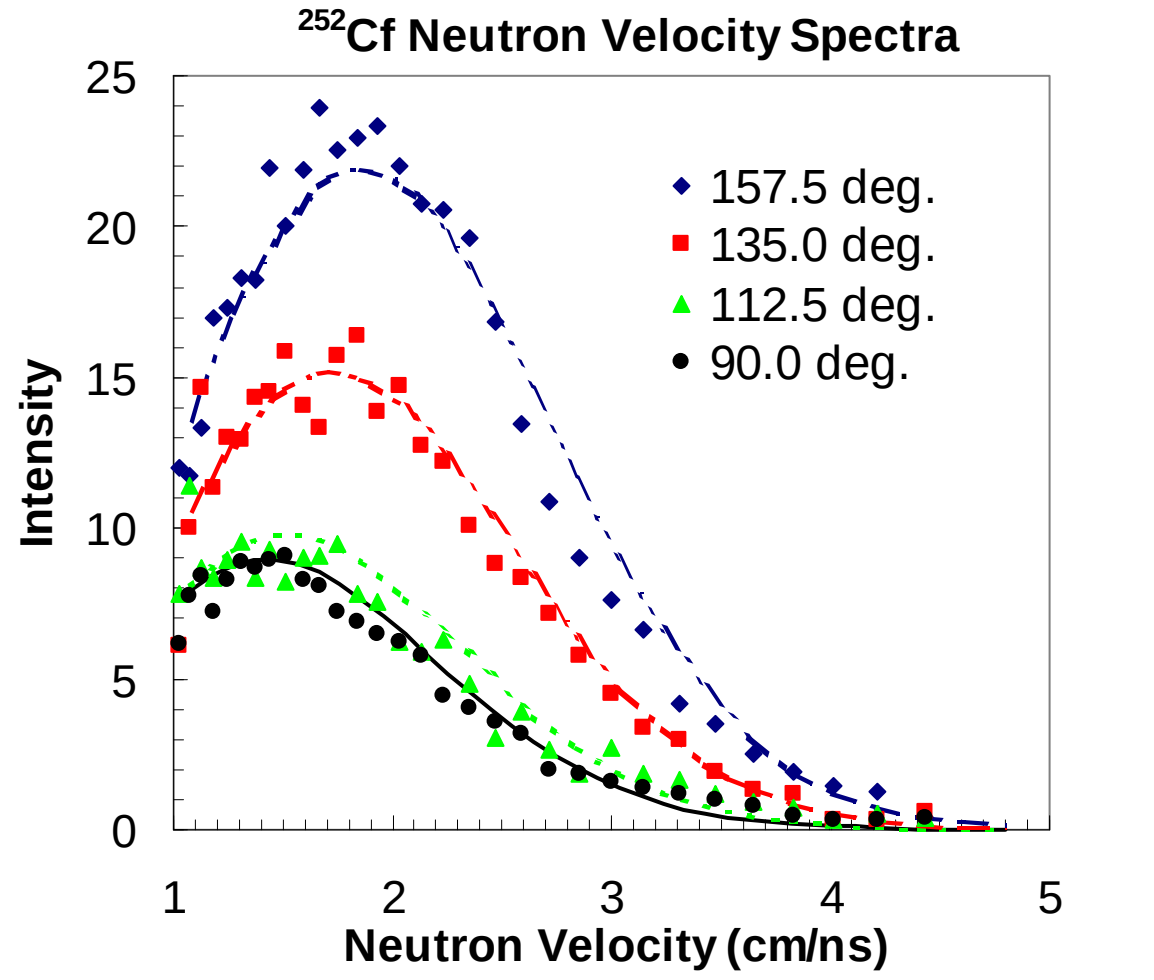

*Figure 12. As for figure 11 but for neutron to light fragment angles $\geq 90°$.*

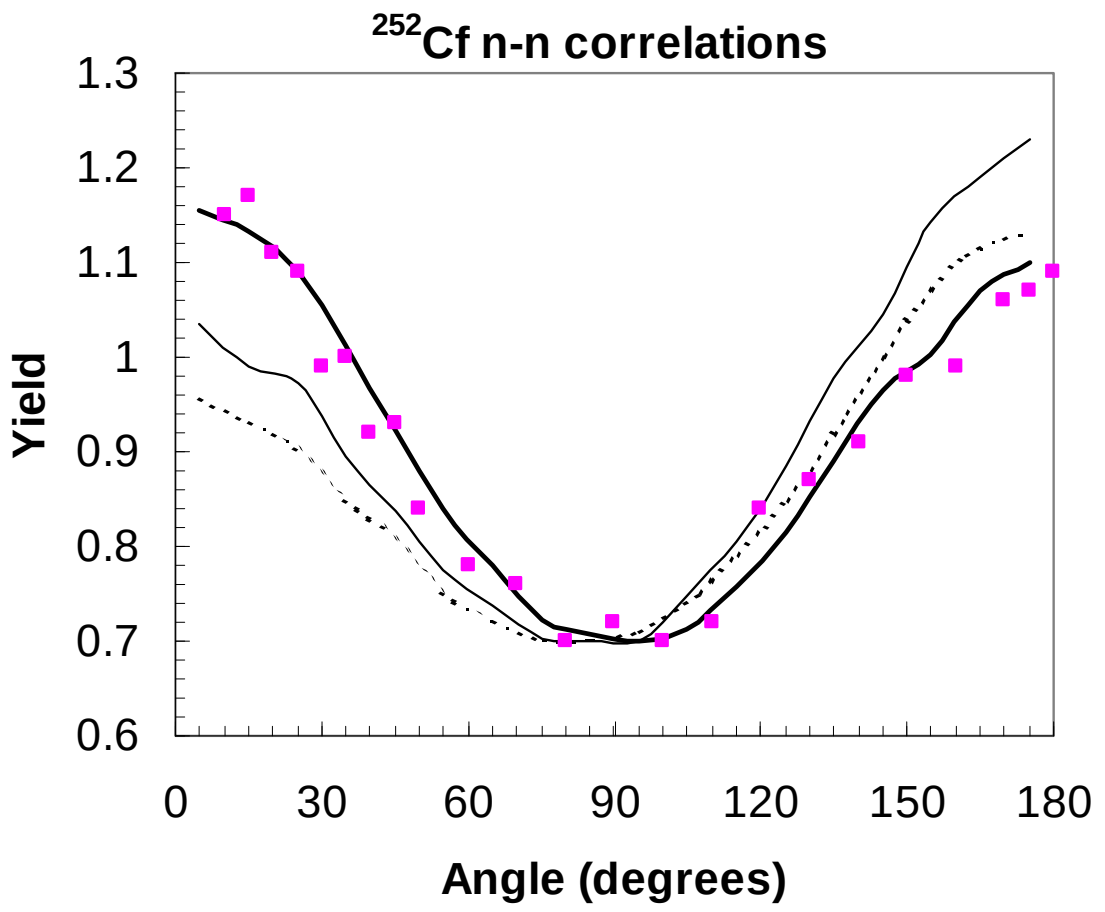

*Figure 13. The squares are the measured $^{252}$Cf neutron-neutron correlation of Pringle and Brooks[7]. The dashed line shows the simulated correlation for all neutron energies with $\sigma_N$=0. The thin solid line shows the corresponding correlation with the experimental neutron detection efficiency of Pringle and Brooks taken into account. The thick solid line shows the simulated correlation with $\sigma_N$=0.75.*

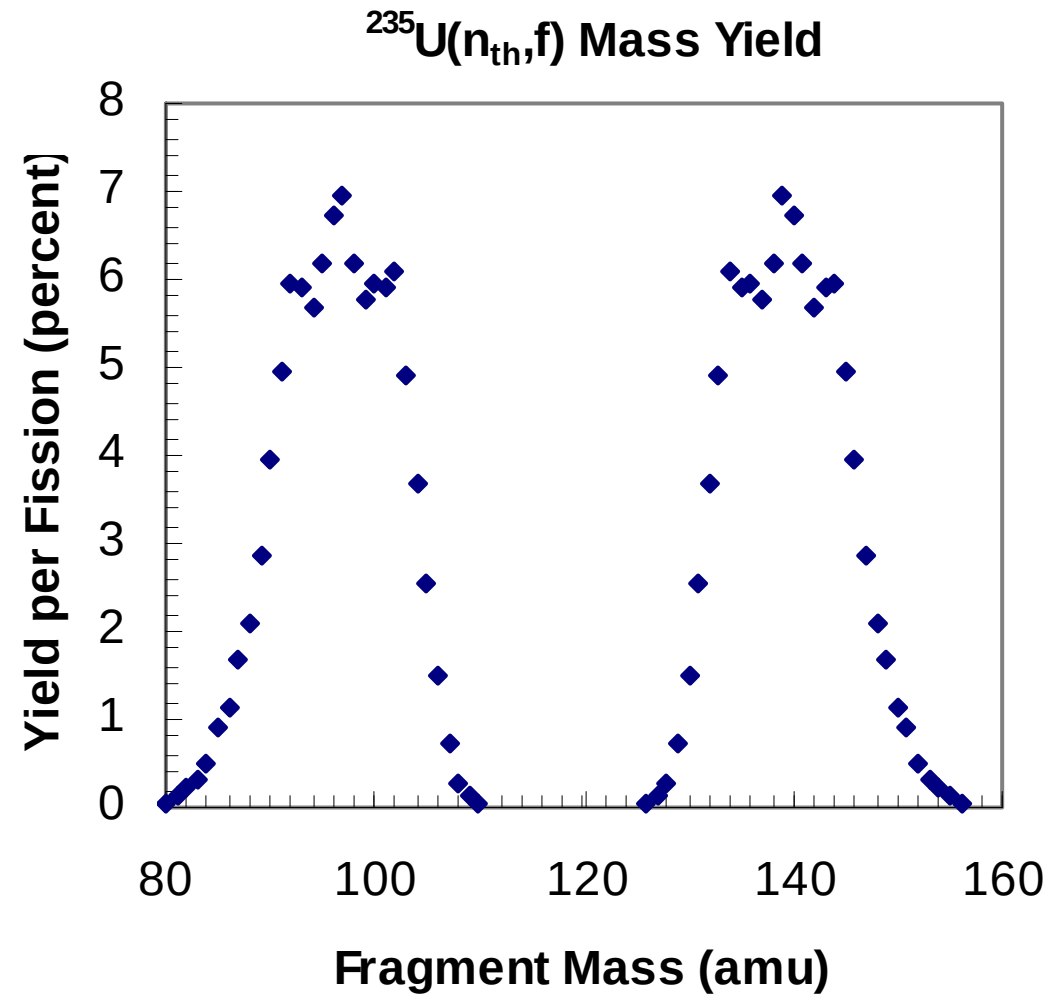

*Figure 14. $^{235}U(n_{th},f)$ mass yield[8].*

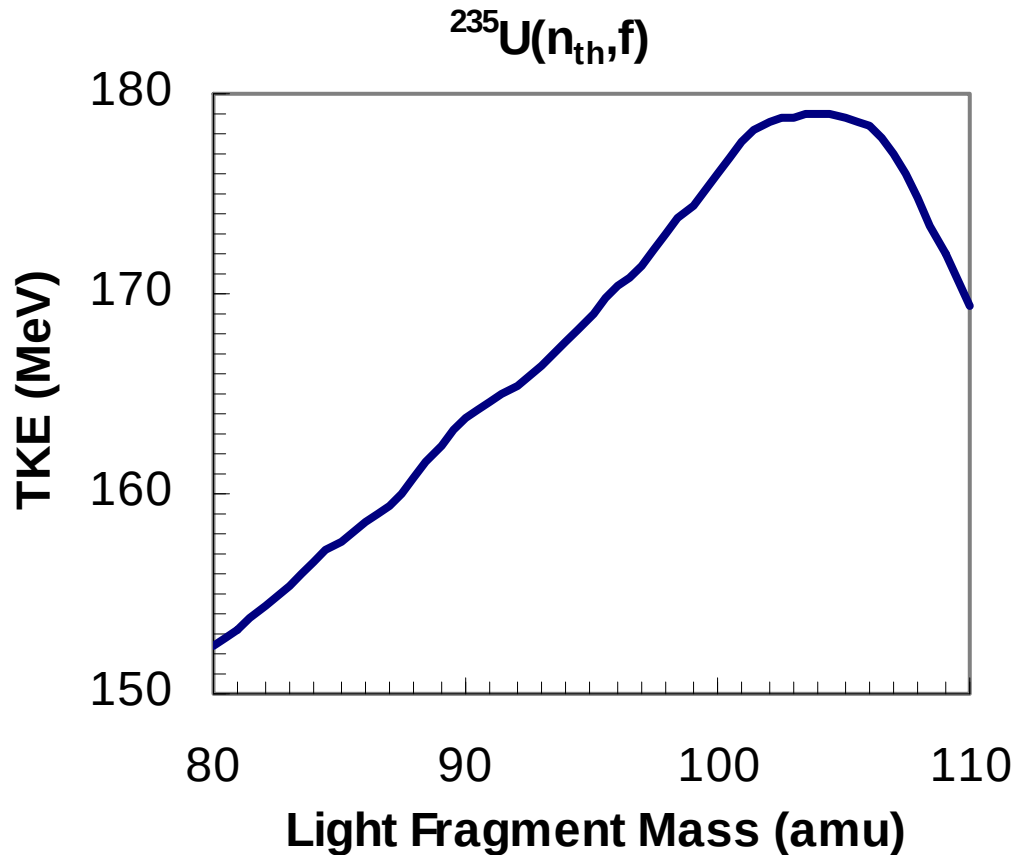

*Figure 15. Pre-neutron emission, TKE release in $^{235}U(n_{th},f)$ as a function of the mass of the light fission fragment[9].*

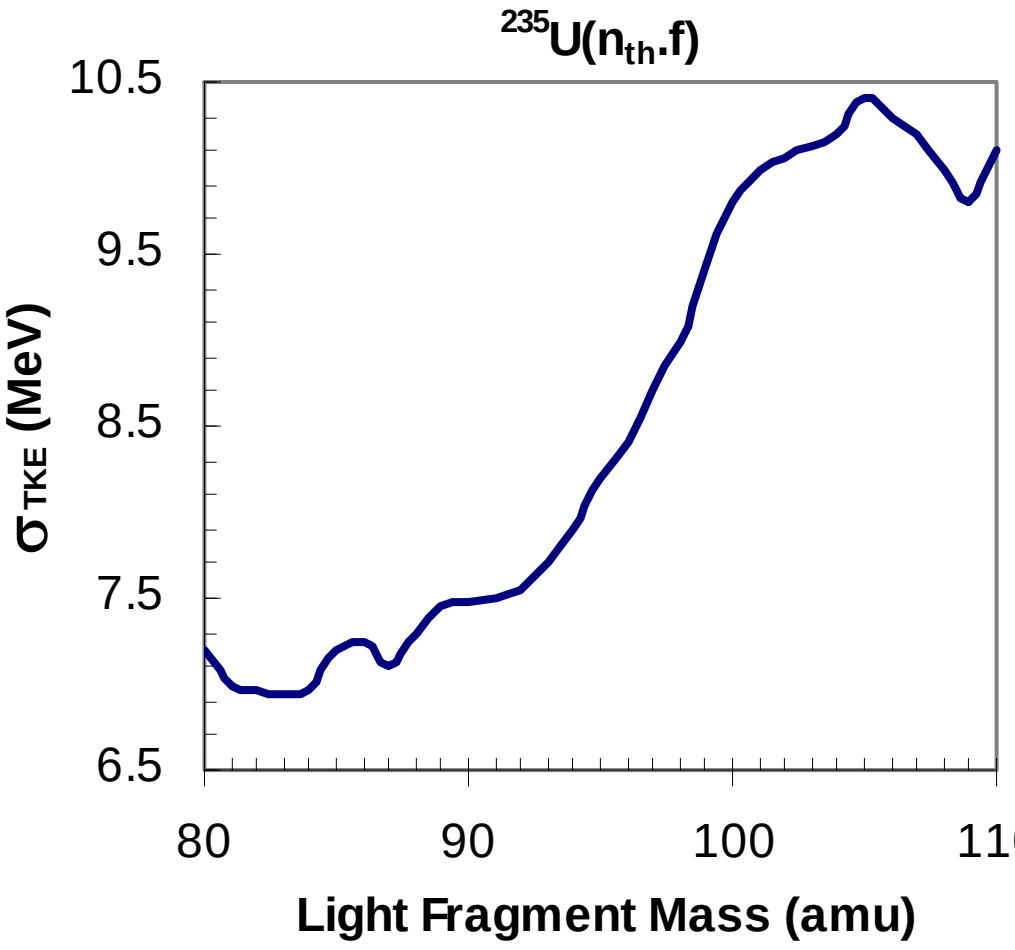

*Figure 16. The standard deviation of the TKE release in $^{235}U(n_{th},f)$ as a function of the mass of the light fission fragment[9].*

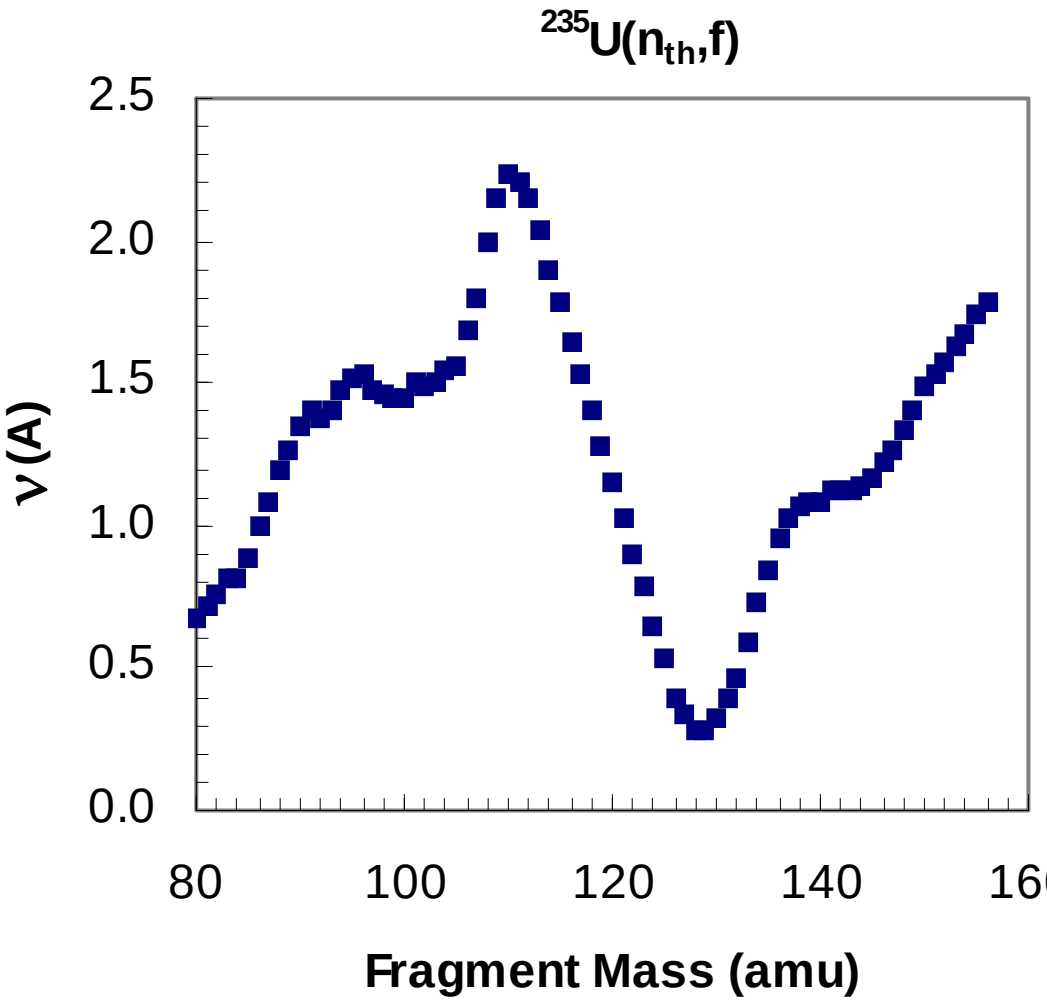

*Figure 17. The neutron multiplicity per fission fragment for $^{235}U(n_{th},f)$[10].*

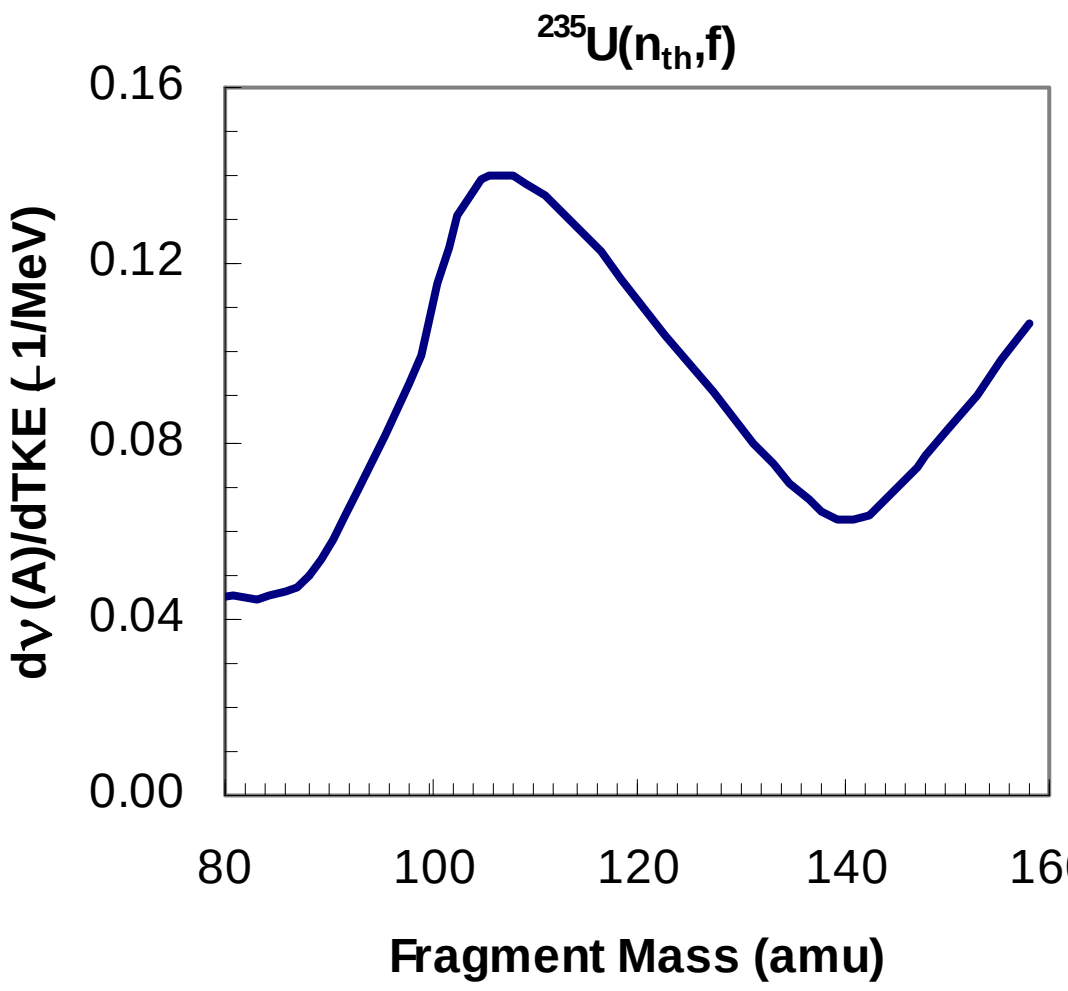

*Figure 18. The dependence of the multiplicity per fragment on the TKE released in $^{235}U(n_{th},f)$. The curve shown here was obtained by shifting the $^{239}Pu(n_{th},f)$ data[11] down by two atomic mass units.*

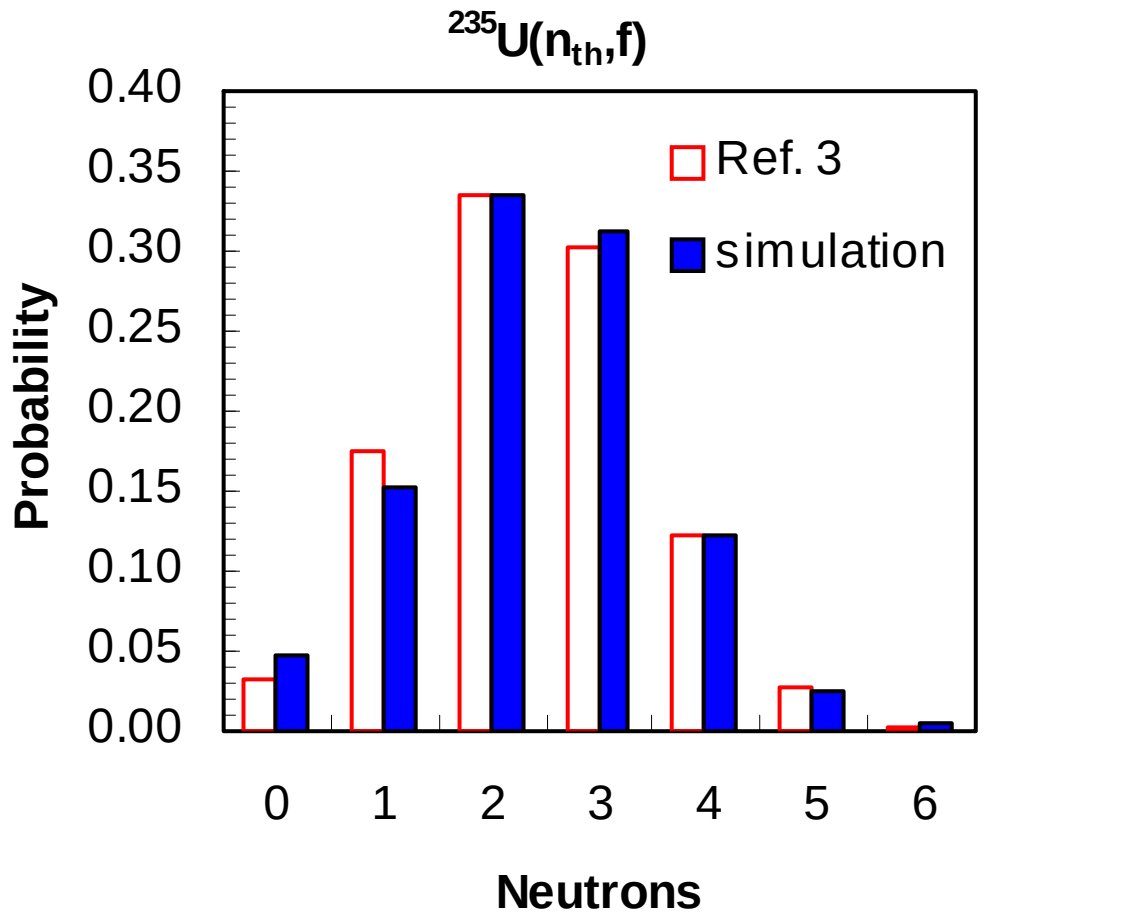

*Figure 19. Comparison of the experimentally deduced $^{235}U(n_{th},f)$ neutron multiplicity distribution of Boldeman and Hines[3] to the corresponding simulated distribution.*

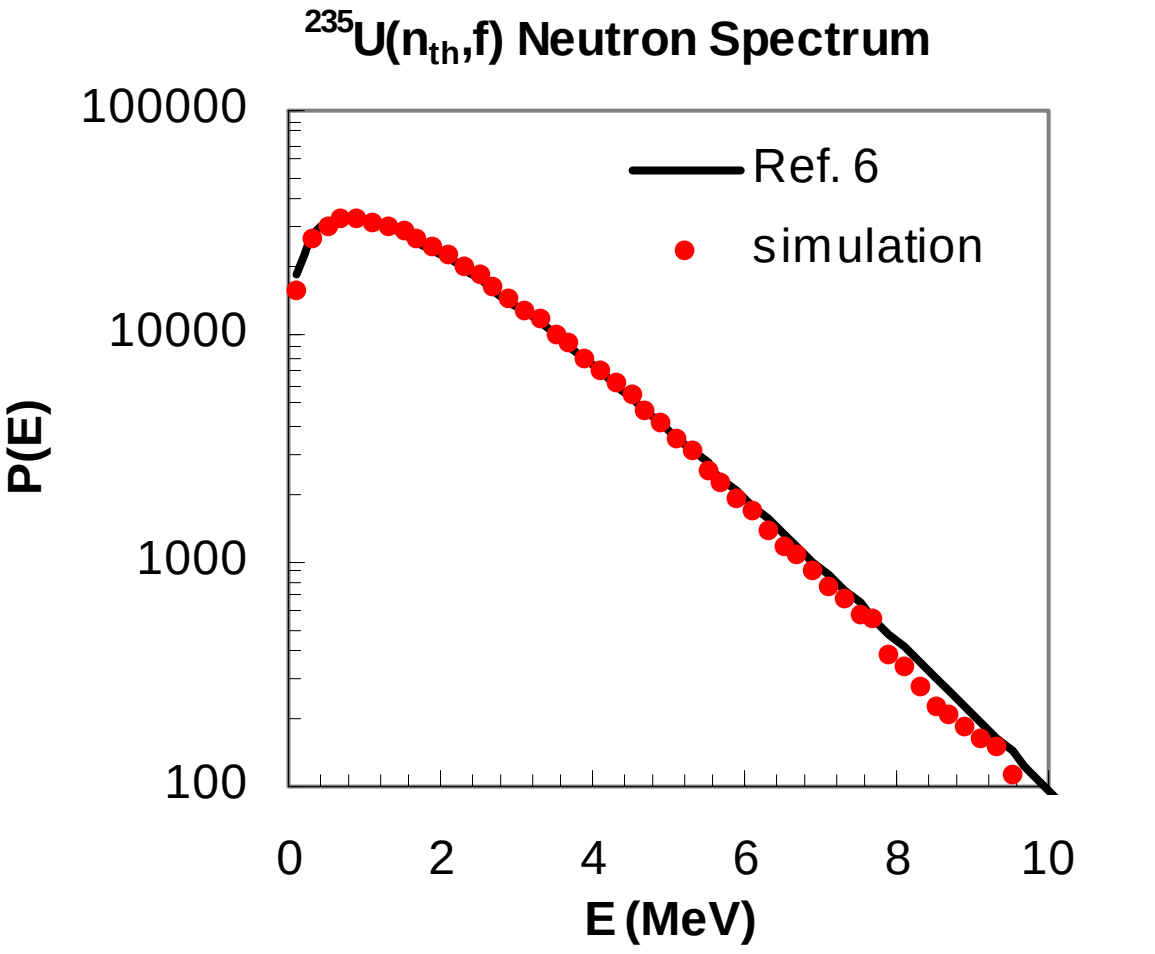

*Figure 20. Comparison of the MCNP4A recommended $^{235}U(n_{th},f)$ neutron-energy spectrum to the corresponding simulated distribution.*

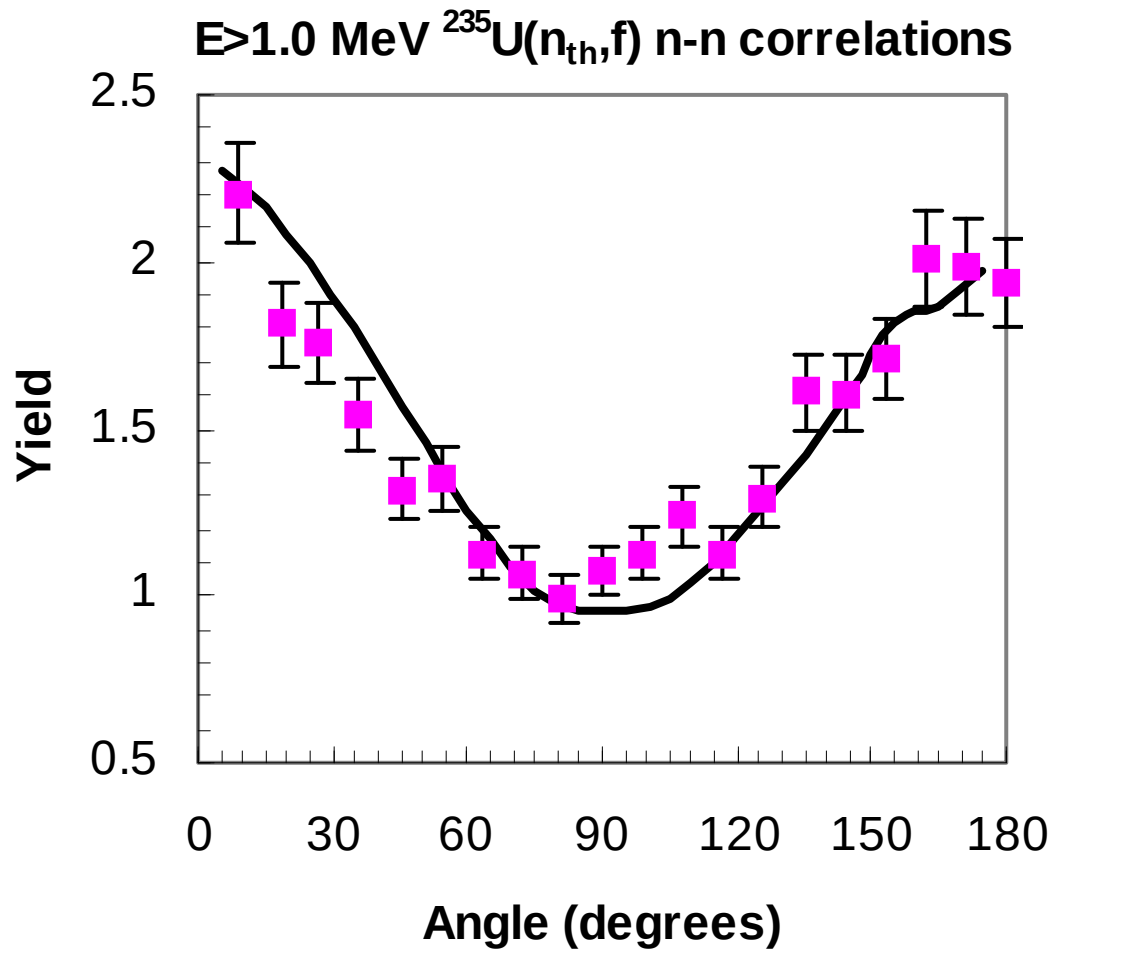

*Figure 21. The squares are the measured E>1.0 MeV $^{235}U(n_{th},f)$ neutron-neutron correlation of Pringle and Brooks[7]. The thick solid line shows the simulated correlation with $\sigma_N$=0.75.*

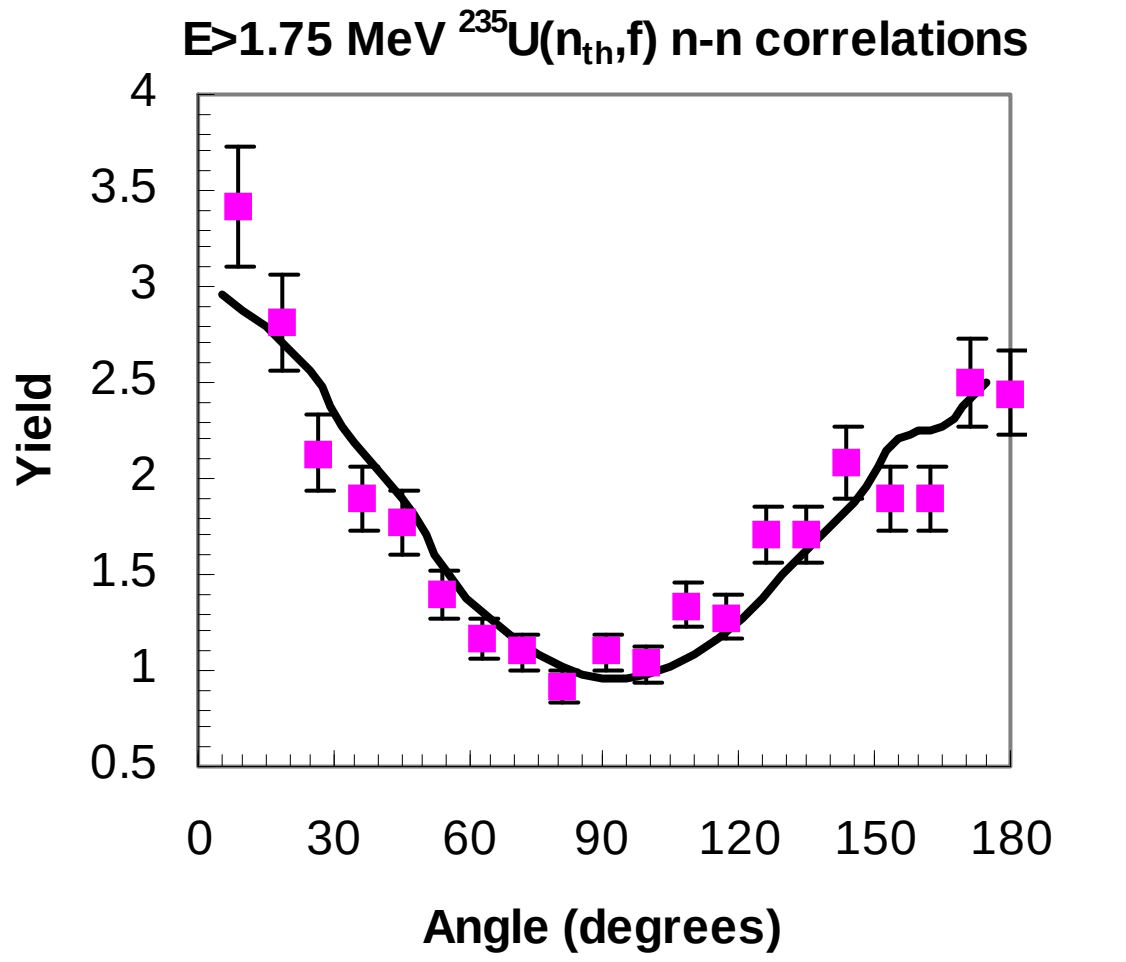

*Figure 22. As for figure 21 but with E>1.75 MeV.*

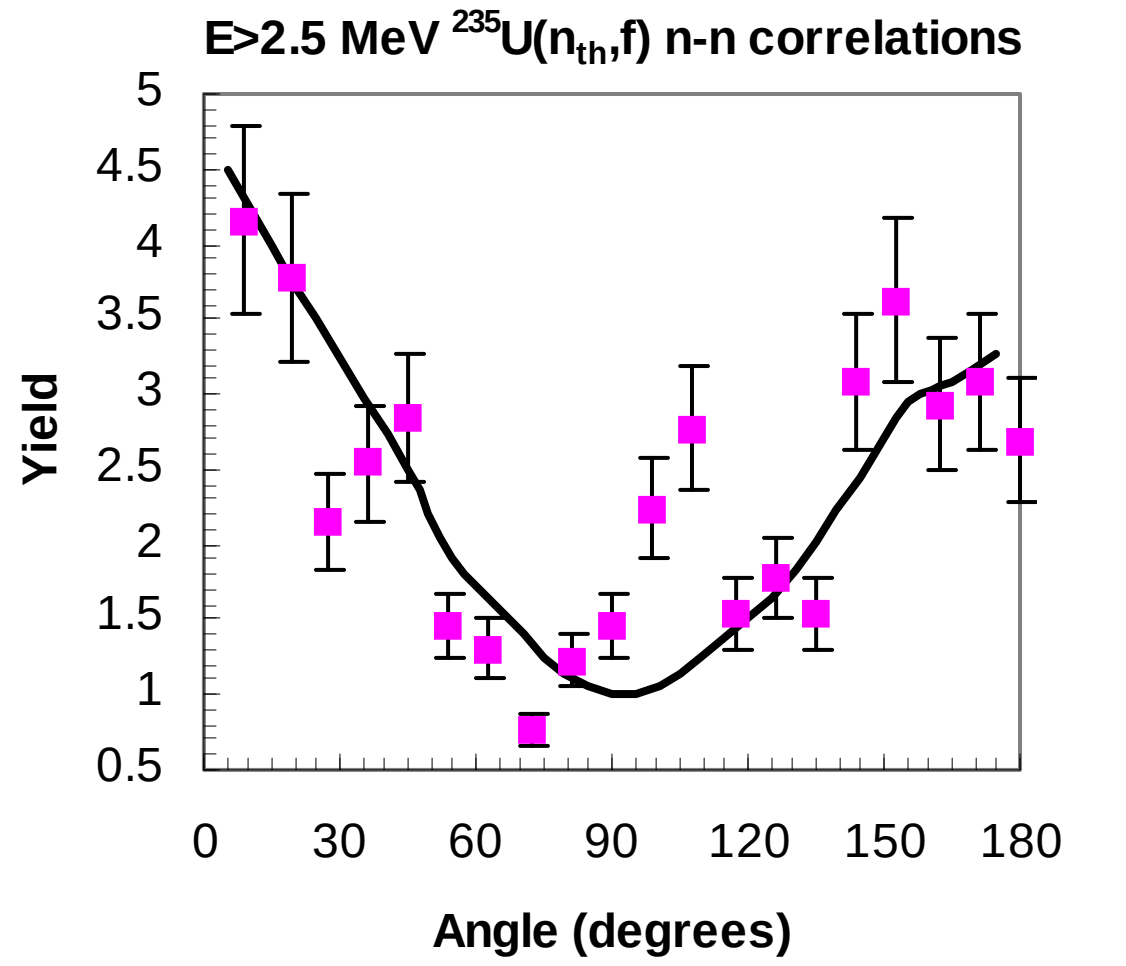

*Figure 23. As for figure 21 but with E>2.5 MeV.*

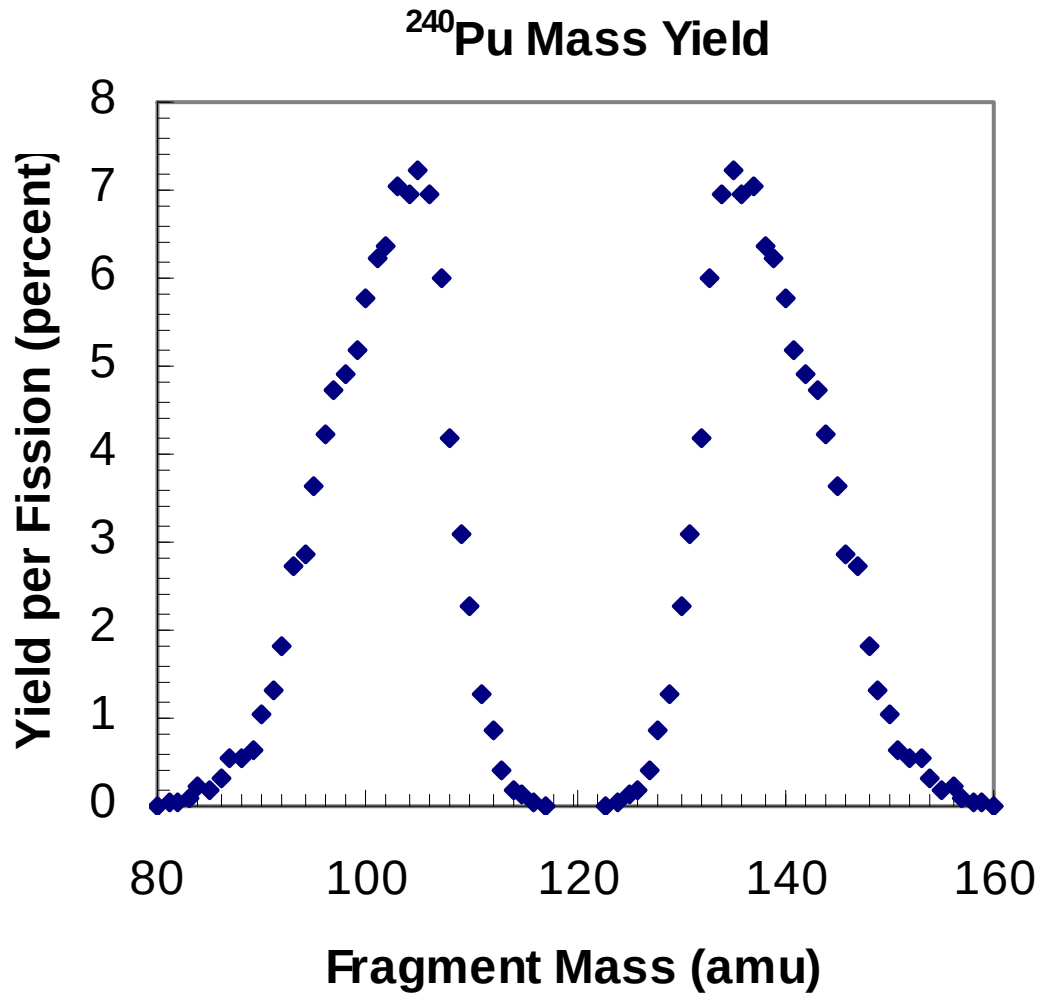

*Figure 24. .* $^{240}$*Pu spontaneous fission mass yield*[13]*.*

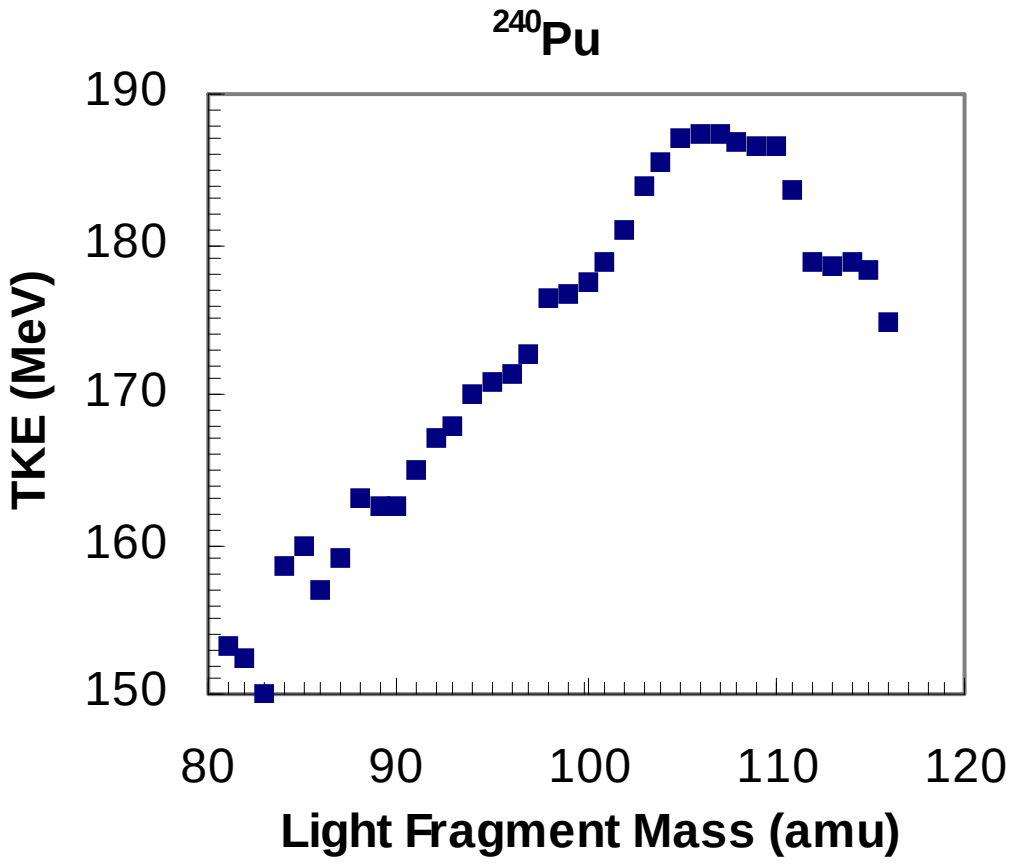

*Figure 25. Pre-neutron emission, TKE release in the spontaneous fission of* $^{240}$*Pu as a function of the mass of the light fission fragment*[13]*.*

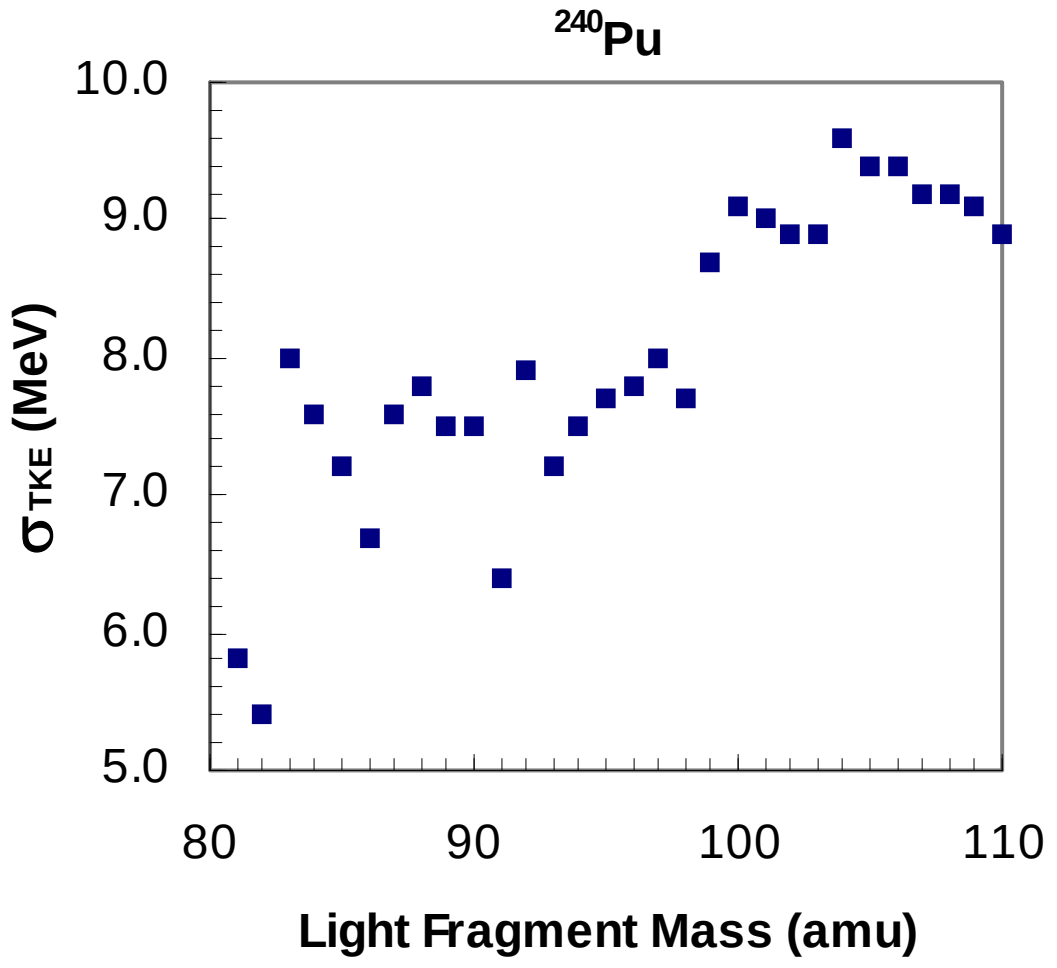

*Figure 26. The standard deviation of the TKE release in spontaneous fission of* $^{240}$*Pu as a function of the mass of the light fission fragment*[13]*.*

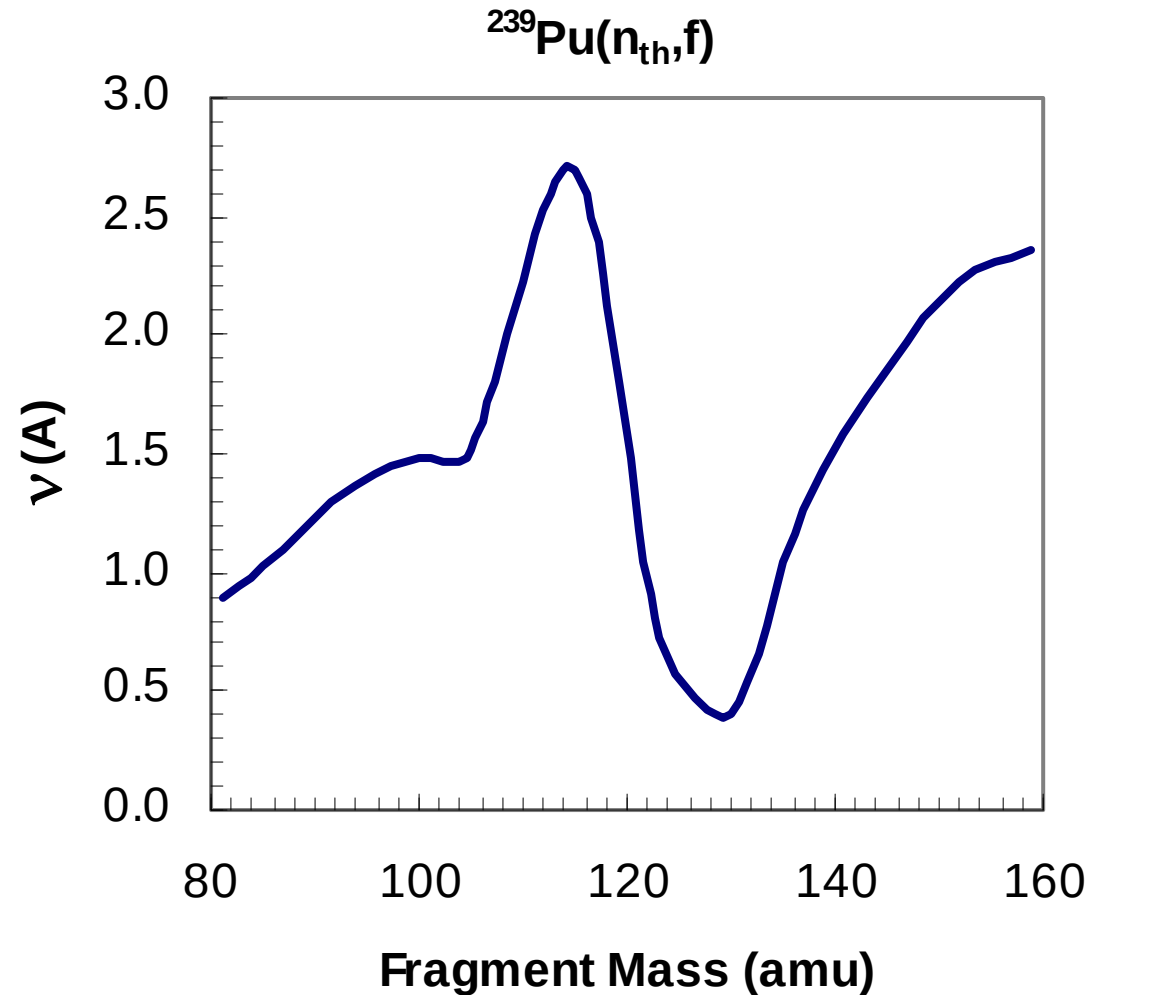

*Figure 27. The neutron multiplicity per fission fragment for* $^{239}Pu(n_{th},f)$[11]*.*

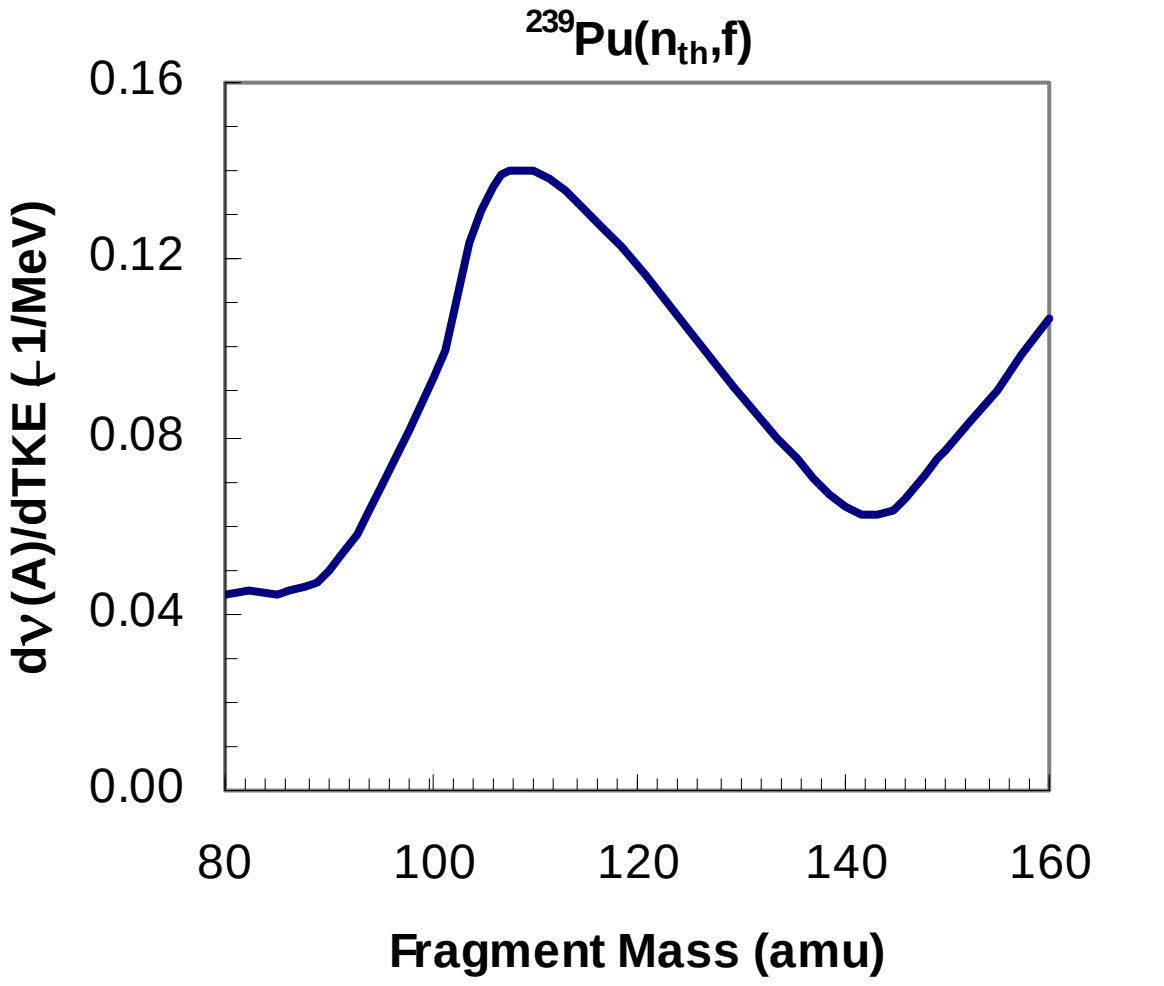

*Figure 28. The dependence of the multiplicity per fragment on the TKE released in* $^{239}Pu(n_{th},f)$[11]*.*

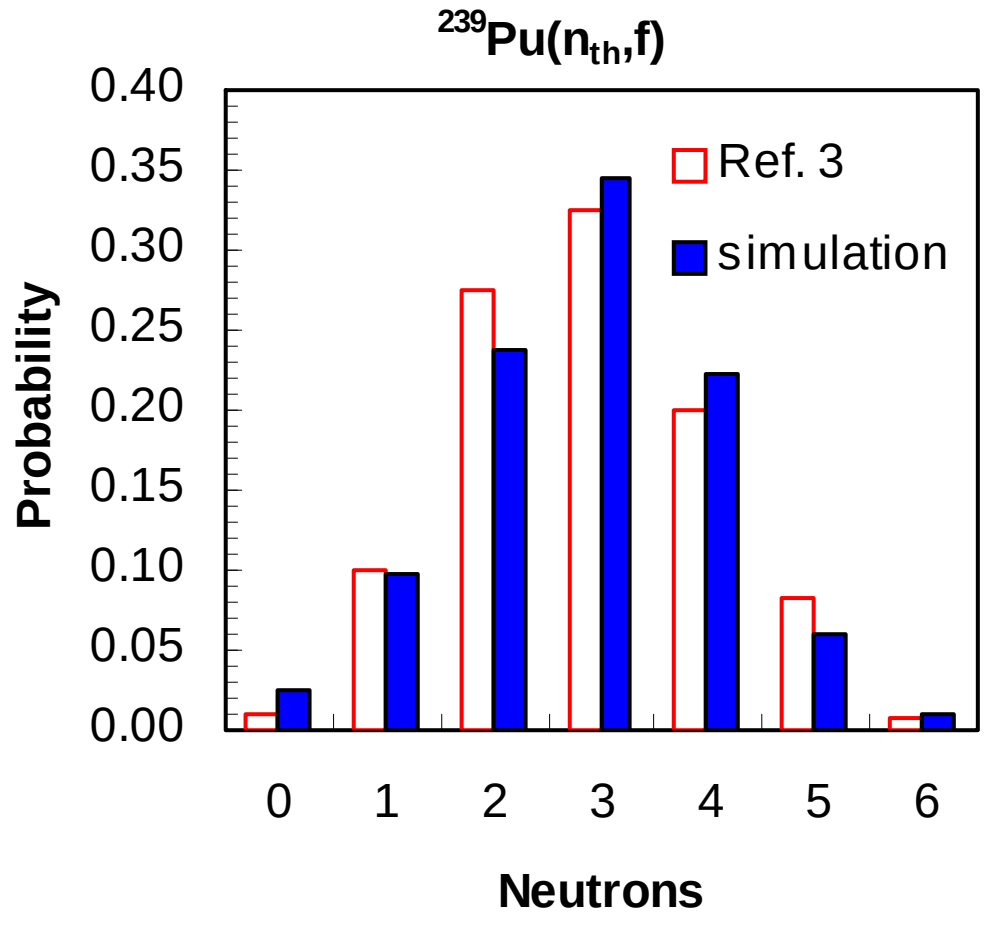

*Figure 29. Comparison of the experimentally deduced* $^{239}Pu(n_{th},f)$ *neutron multiplicity distribution of Boldeman and Hines*[3] *to the corresponding simulated distribution.*

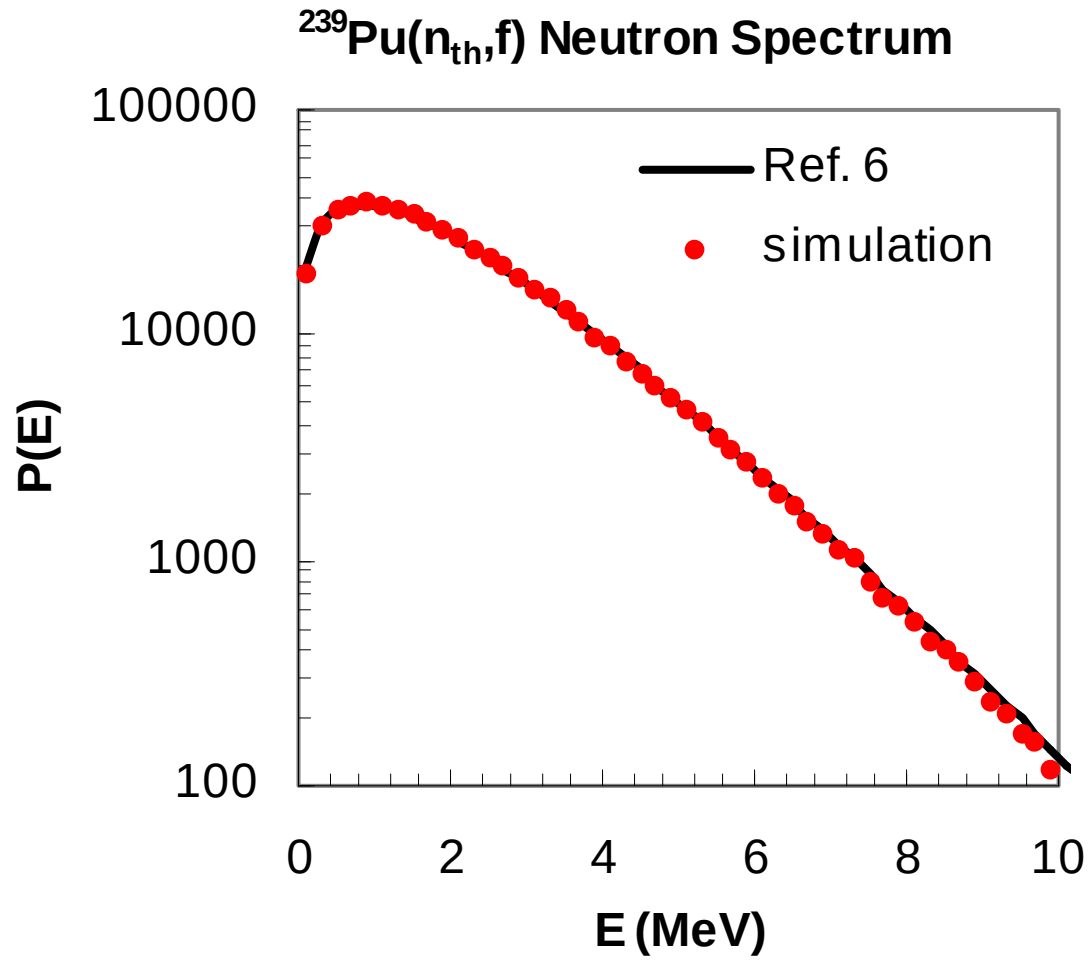

*Figure 30. Comparison of the MCNP4A recommended* $^{239}$*Pu(n*$_{th}$*,f) neutron-energy spectrum to the corresponding simulated distribution.*

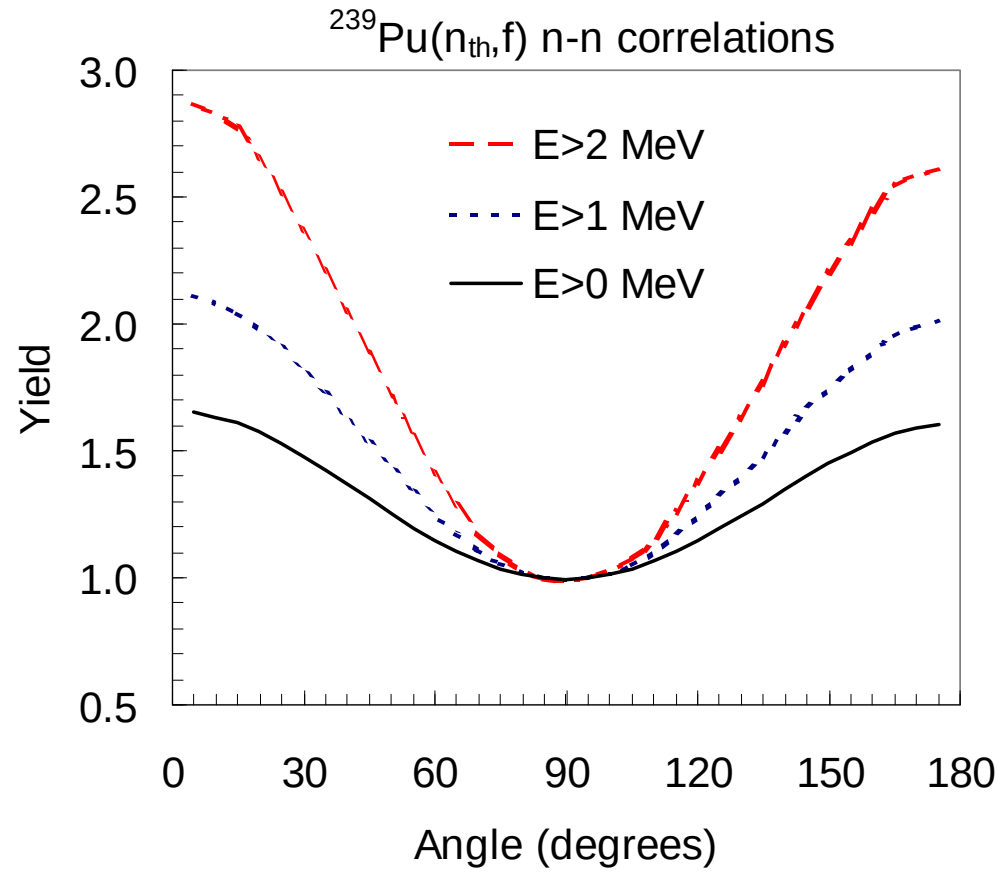

*Figure 31. Simulated* $^{239}$*Pu(n*$_{th}$*,f) neutron-neutron correlations with various neutron energy thresholds.*

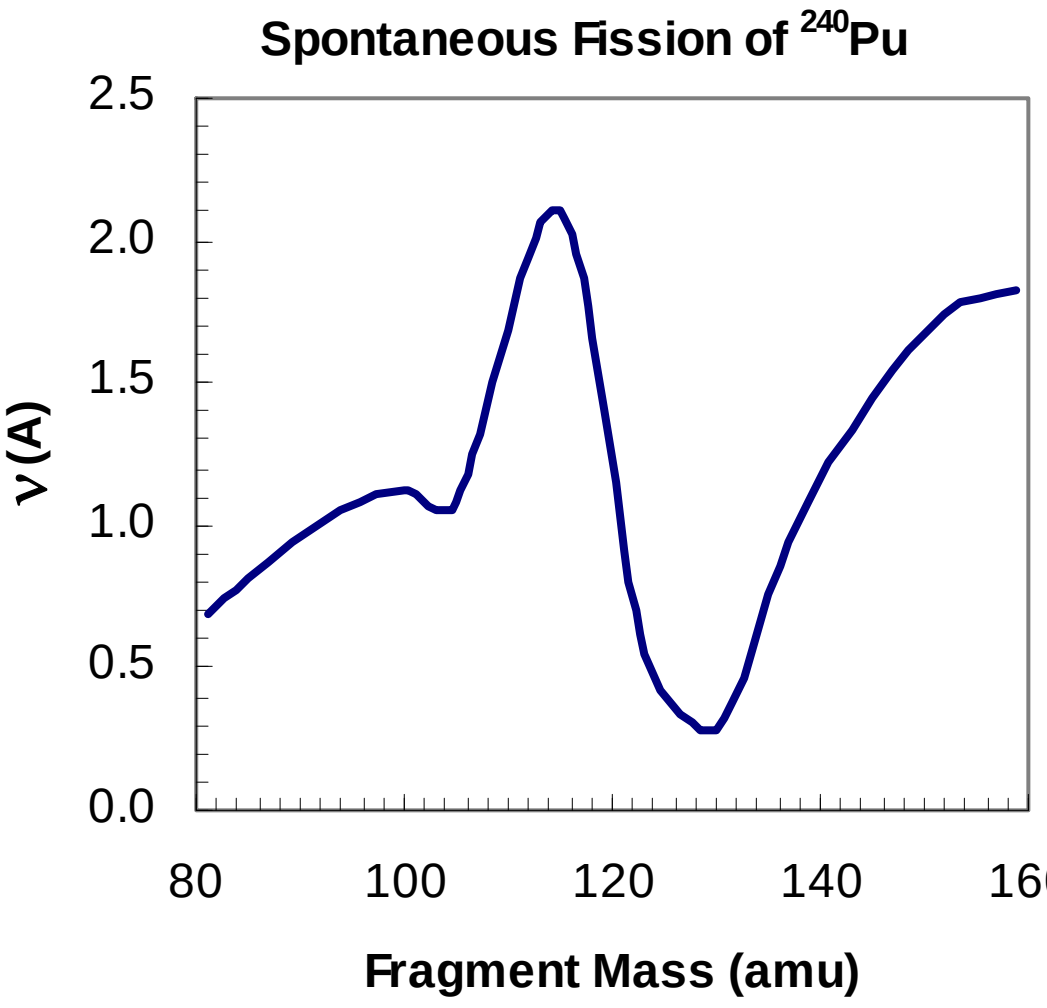

*Figure 32. The neutron multiplicity per fission fragment for the spontaneous fission of* $^{240}$*Pu (see text).*

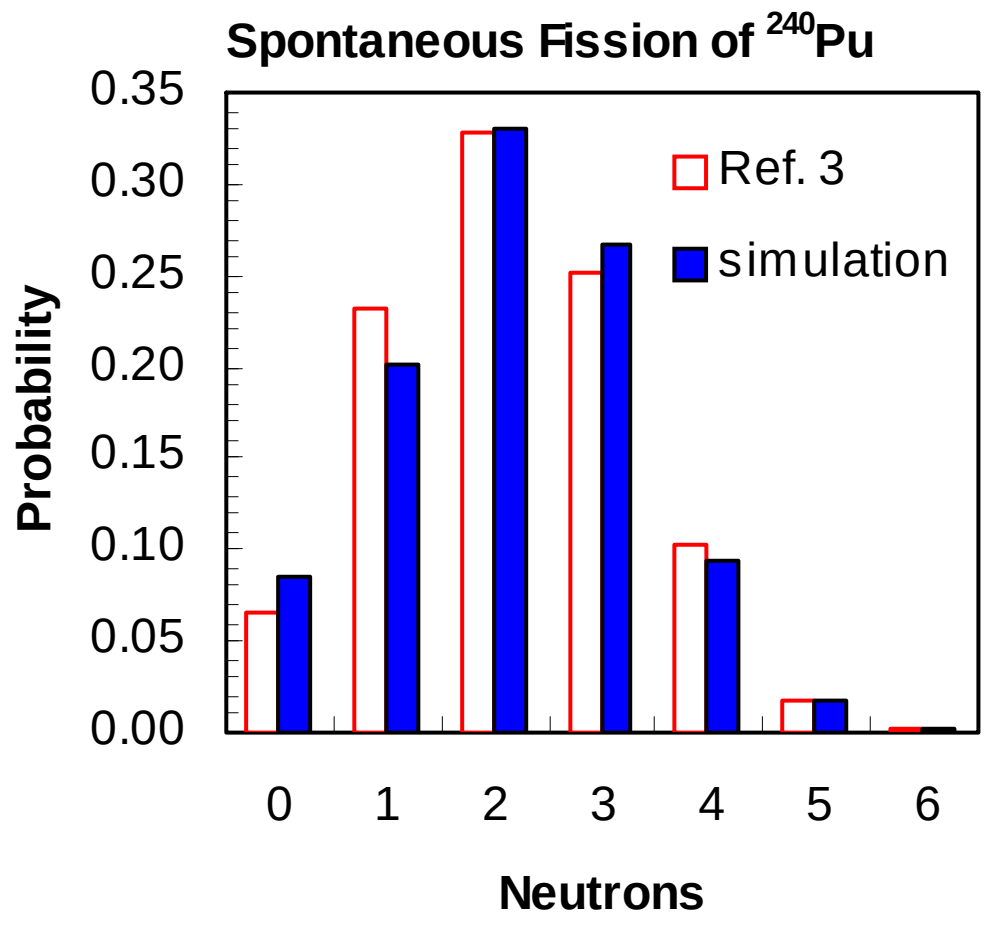

*Figure 33. Comparison of the experimentally deduced* $^{240}$*Pu spontaneous fission neutron multiplicity distribution of Boldeman and Hines*[3] *to the corresponding simulated distribution.*

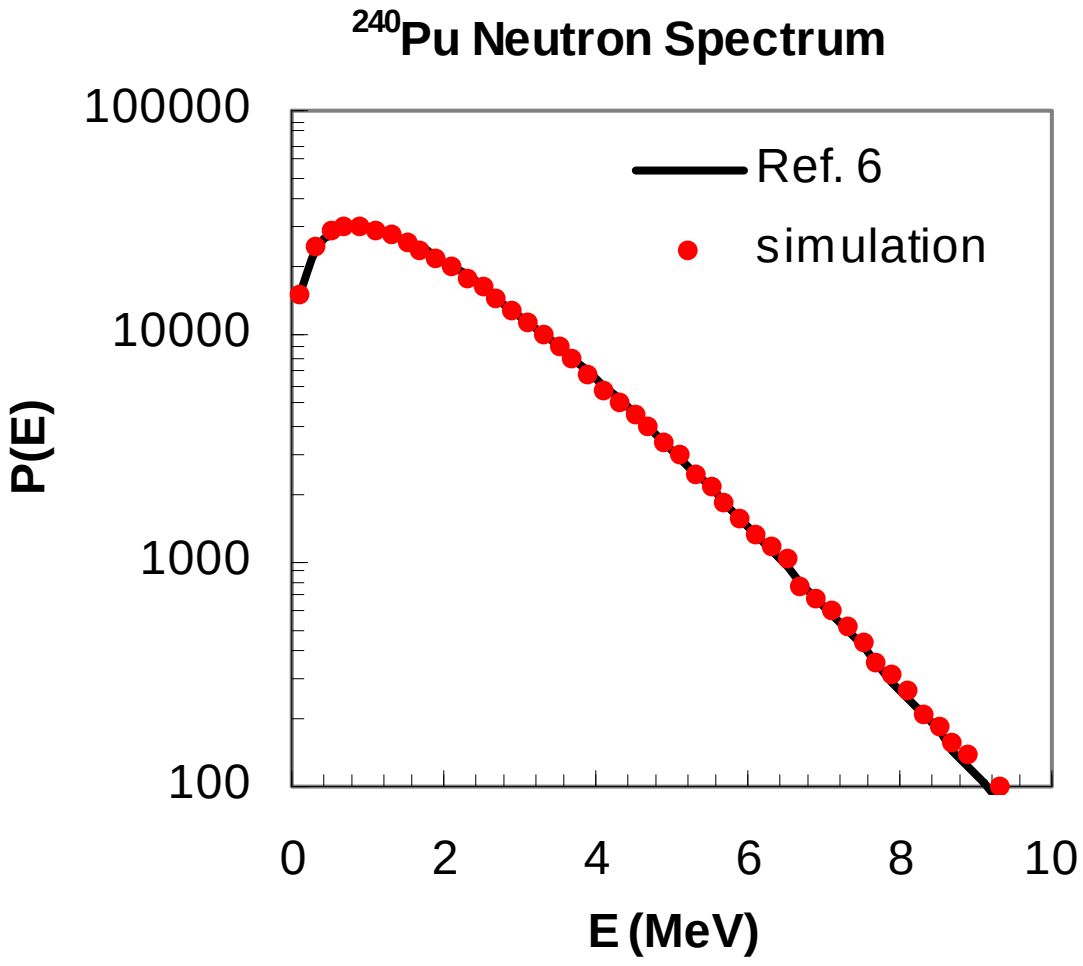

*Figure 34. Comparison of the MCNP4A recommended* $^{240}$*Pu spontaneous fission neutron-energy spectrum to the corresponding simulated distribution.*

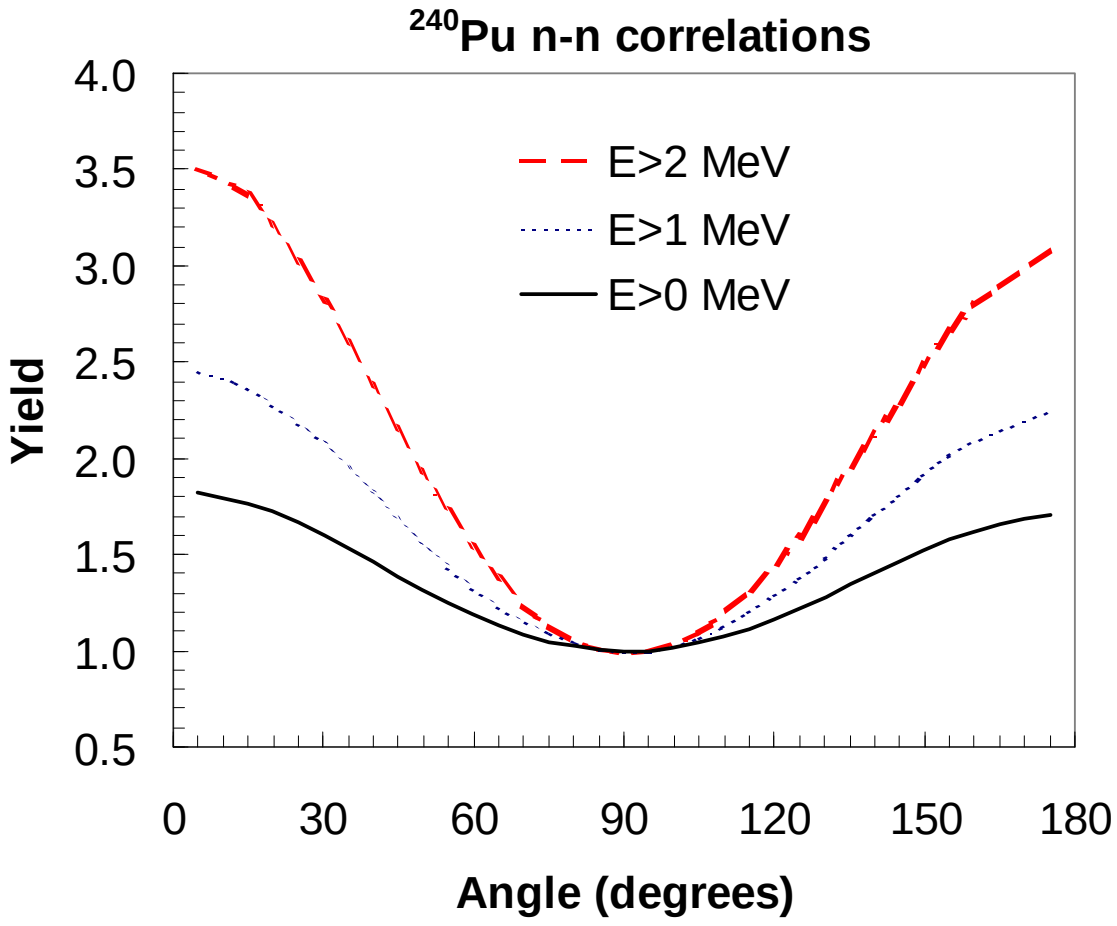

*Figure 35. Simulated* $^{240}$*Pu spontaneous fission neutron-neutron correlations with various neutron energy thresholds.*